\documentclass{aa}  
\usepackage{graphicx}
\usepackage{txfonts}
\usepackage{longtable}
\begin{document} 

   \title{The GAPS Programme with HARPS-N at TNG \\
{\Large   XVII. Line profile indicators and kernel regression as diagnostics of radial-velocity variations due to stellar activity in solar-like stars}\thanks{Based on observations made with the Italian Telescopio Nazionale Galileo (TNG) operated on the island of La Palma by the Fundaci\'on Galileo Galilei of the INAF (Istituto Nazionale di Astrofisica) at the Spanish Observatorio del Roque de los Muchachos of the Instituto de Astrofisica de Canarias.}\fnmsep\thanks{The IDL macro to compute the line profile indicators is available in electronic form
at the CDS via anonymous ftp to cdsarc.u-strasbg.fr (130.79.128.5) or via http://cdsweb.u-strasbg.fr/cgi-bin/qcat?J/A+A/}}
   \authorrunning{Lanza et al.}
    \titlerunning{Line profile indicators as activity diagnostics}
   \subtitle{}
\author{
A.~F.~Lanza,\inst{1}
L.~Malavolta,\inst{2,3} 
S.~Benatti,\inst{3} 
S.~Desidera,\inst{3}  
A.~Bignamini,\inst{4}
A.~S.~Bonomo,\inst{5}  
M.~Esposito,\inst{6}
P.~Figueira,\inst{7,8}
R.~Gratton,\inst{3}
G.~Scandariato,\inst{1}
M.~Damasso,\inst{5} 
A.~Sozzetti,\inst{5}
K.~Biazzo,\inst{1} 
R.~U.~Claudi,\inst{3}  
 R.~Cosentino,\inst{10} E.~Covino, \inst{6} A.~Maggio, \inst{9} S.~Masiero, \inst{9} G.~Micela, \inst{9} E.~Molinari,\inst{10,11} 
 I.~Pagano, \inst{1} G.~Piotto, \inst{2,3} E.~Poretti,\inst{10,12}  
R.~Smareglia,\inst{4} 
L.~Affer,\inst{9} 
C.~Boccato,\inst{3}
F.~Borsa,\inst{12}
W.~Boschin,\inst{10,13,14}
P.~Giacobbe,\inst{5}
C.~Knapic, \inst{4}
G.~Leto, \inst{1}
J.~Maldonado,\inst{9}
L.~Mancini, \inst{15,16,5}
A.~Martinez~Fiorenzano,\inst{10}
S.~Messina,\inst{1}
V.~Nascimbeni,\inst{2}
M.~Pedani,\inst{10}
M.~Rainer\inst{12,17}
}
\institute{INAF -- Osservatorio Astrofisico di Catania, Via S.~Sofia 78,  I-95123 Catania, Italy\\
              \email{nuccio.lanza@oact.inaf.it} 
\and Dipartimento di Fisica e Astronomia Galileo Galilei -- Universit\`a di Padova, Via Francesco Marzolo 8,  I-35122, Padova, Italy    
\and INAF -- Osservatorio Astronomico di Padova,  Vicolo dell'Osservatorio 5, I-35122, Padova, Italy 
\and INAF -- Osservatorio Astronomico di Trieste, via Tiepolo 11, I-34143 Trieste, Italy  
\and  INAF -- Osservatorio Astrofisico di Torino, Via Osservatorio 20, I-10025, Pino Torinese, Italy  
\and INAF -- Osservatorio Astronomico di Capodimonte, Salita Moiariello 16, I-80131, Napoli, Italy  
\and European Southern Observatory,  Alonso de Cordova 3107, Vitacura, Santiago, Chile
\and Instituto de Astrof\'{i}sica e Ci\^{e}ncias do Espa\c{c}o, Universidade do Porto, CAUP, Rua das Estrelas, PT4150-762 Porto, Portugal 
\and INAF -- Osservatorio Astronomico di Palermo, Piazza del Parlamento 1, I-90134, Palermo, Italy 
\and Fundaci\'on Galileo Galilei - INAF, Rambla Jos\'e Ana Fernandez P\'erez 7, E-38712 Bre\~na Baja, TF, Spain 
\and INAF -- Osservatorio Astronomico di Cagliari, Via della Scienza 5, I-09047 Selargius (CA) , Italy
\and INAF -- Osservatorio Astronomico di Brera, Via E. Bianchi 46, I-23807 Merate (LC), Italy  
\and Instituto de Astrofisica de Canarias, C/Via Lactea s/n, E-38205 La Laguna, TF, Spain  
\and Dep. de Astrofisica, Univ. de La Laguna, Av. del Astrofisico Francisco Sanchez s/n, E-38205 La Laguna, TF, Spain
\and Dipartimento di Fisica, Universit\`a di Roma "Tor Vergata", Via della Ricerca Scientifica 1,  I-00133 Roma, Italy 
\and Max Planck Institute for Astronomy, K\"{o}nigstuhl 17, 69117 Heidelberg, Germany
\and INAF -- Osservatorio Astrofisico di Arcetri, Largo E.~Fermi 5, I-50125 Firenze, Italy
}

   \date{Received ... ; accepted ...}

\abstract{}{Stellar activity is the ultimate source of radial-velocity (hereinafter RV) noise in the search for Earth-mass planets orbiting late-type main-sequence stars. We analyse the performance of four different indicators and the chromospheric index $\log R^{\prime}_{\rm HK}$ in detecting RV variations induced by stellar activity in 15 slowly rotating ($v \sin i \leq 5 $~km\,s$^{-1}$), weakly active ($\log R^{\prime}_{\rm HK} \leq -4.95$) solar-like stars observed with the high-resolution spectrograph High Accuracy Radial velocity Planet Searcher for the Northern hemisphere (HARPS-N).}{We consider indicators of the asymmetry of the cross-correlation function (CCF) between the stellar spectrum and the binary weighted line mask used to compute the RV, that is the bisector inverse span (BIS), $\Delta V$, and a new indicator $V_{\rm asy (mod)}$ together with the full width at half maximum (FWHM) of the CCF. We present methods to evaluate the uncertainties of the CCF indicators and apply a  kernel regression (KR) between the RV, the time, and each of the indicators to study their capability of reproducing the RV variations induced by stellar activity.}{The considered  indicators together with the KR prove to be useful to detect activity-induced RV variations in $\sim 47 \pm 18$~percent of the stars over a two-year time span when a significance (two-sided p-value) threshold of one percent is adopted. In those cases, KR reduces the standard deviation of the RV time series by a factor of approximately two. The BIS, the FWHM, and the newly introduced $V_{\rm asy(mod)}$ are the best indicators, being useful in $27 \pm 13$, $13 \pm 9$,  and $13 \pm 9$~percent of the cases, respectively.  The relatively limited performances of the activity indicators are related to the very low activity level and $v \sin i$  of the considered stars. For the application of our approach to sun-like stars,  a spectral resolution  allowing $\lambda/\Delta \lambda \geq 10^{5}$ and highly stabilized spectrographs are recommended.}{}
  \maketitle
%

\section{Introduction}
The search for rocky planets around late-type stars has pushed the measurement of stellar radial velocity (hereafter RV) into the m\,s$^{-1}$ regime because of the very low mass ratio of those planets relative to their host stars \citep[e.g.][]{Mayoretal09,Quelozetal09,Pepeetal11,Diazetal16}. At that level of precision, even the most stable stars show intrinsic RV variations on a variety of timescales, ranging from a few minutes, characteristic of p-mode oscillations, to decades,  associated with activity cycles produced by the modulation of their surface magnetic fields. 

Variations on timescales from minutes to hours, induced by oscillations and photospheric convection (granulation), can be significantly reduced by averaging RV measurements collected with a suitable cadence \citep{Dumusqueetal11}. 
The effects of photospheric active regions (hereafter ARs), with a lifespan comparable with the stellar rotation period or longer, are much more subtle and difficult to identify and remove. The magnetic fields of ARs produce a perturbation of the local brightness that induces distortions on the profiles of photospheric lines (the so-called flux effect) and an attenuation of the local convective  blueshifts of the spectral lines, which manifests itself as an  apparent redshift in the disc-integrated RV  \citep[see][and references therein]{Lagrangeetal10,Meunieretal10,Lanzaetal11}. The flux effect is roughly proportional to the filling factor of  the ARs and the projected rotation velocity of the star, that is its $v\sin i$ \citep{Desortetal07}, while its sign depends on the position on the stellar disc and the contrast of the brightness inhomogeneities,  being opposite for cool spots and bright faculae. On the other hand, the convective shift effect has always the same sign for both kind of inhomogeneities. 

In the Sun,  photospheric faculae have a total area approximately one order of magnitude larger than sunspots, {while their contrast is very low at the disc centre and increases towards the limb \citep[cf.][]{Unruhetal99}}. Together with the small $v\sin i$, this makes the convective shift effect the dominant RV perturbation in the models computed by \citet{Lagrangeetal10} and \citet{Meunieretal10}. Their predictions have been confirmed by observations on timescales ranging from a few rotation periods to the eleven-year activity cycle. The amplitude of the RV variation of the Sun as a star can reach up to $\sim 10-12$ m\,s$^{-1}$ on timescales significantly shorter than the eleven-year cycle, in some cases as short as a few tens or hundreds of days \citep{Haywoodetal16,Lanzaetal16}. On a timescale of one week, an rms of 1.33 m$\,$s$^{-1}$ has been measured  with the High Accuracy Radial velocity Planet Searcher for the Northern hemisphere \citep[HARPS-N;][]{Dumusqueetal15}. Those amplitudes are comparable to or larger than those expected from an Earth-mass planet {orbiting an F, G, or K-type star with a period of days}, thus making stellar activity the most challenging source of false positives in the detection of telluric planets around solar-like stars \citep{Fischeretal16}. 

Several approaches have been proposed to mitigate the effects of activity on RV time series. They are particularly useful in the case of late-type stars whose ARs are stable for at least a few consecutive rotations, thus allowing us to constrain their longitudes and parameters from spectroscopic or simultaneous photometric data \citep[e.g. ][]{Boisseetal09, Lanzaetal11,Haywoodetal14,Donatietal14}, while they are of limited use for stars whose AR lifespan is shorter than  the rotation period {as happens in the Sun.} 

In this work, we shall focus on the use of indicators of line profile asymmetries as can be derived from the cross-correlation function (CCF) between the stellar spectrum and a binary weighted line mask, specifically that used to measure the RV itself {(cf. Sect~\ref{observations} for details on the computation of the CCF)}. All the RV information content of the spectrum is contained in the CCF. Fiber-fed stabilized spectrographs, such as HARPS or the 
Spectrographe pour l'Observation des Ph\'enom\`enes des Int\'erieurs stellaires et des Exoplan\`etes (SOPHIE), have a very stable wavelength reference frame and provide a CCF with a signal-to-noise ratio typically  $\geq 10^{3}$  {\citep{Udryetal06}}. The subtle line profile asymmetries induced by stellar ARs in the case of a moderately rotating star  ($v \sin i \la 10-15$~km~s$^{-1}$) cannot be identified in the individual lines, while they can be detected in the high signal-to-noise CCF. 

The first indicator of line profile asymmetry was the line bisector \citep{Quelozetal01}, which is a sensitive indicator in the case of rapidly rotating stars whose variability is dominated by cool spots. The amplitude of its variation is approximately proportional to the spot filling factor and the square of {projected rotational velocity} $v\sin i$ \citep[cf.][and references therein]{Hebrardetal14}. Therefore, for slowly rotating stars ($v \sin i \la 5-7$~km~s$^{-1}$), other indicators have been introduced in an attempt to improve sensitivity to activity-induced variations \citep[e.g.][]{Boisseetal11,Figueiraetal13}. The main advantage over chromospheric indicators is that asymmetry indicators measure the variation of the same CCF profile used to derive the stellar RV. A CCF can be regarded as  an average photospheric line which, thanks to its very high signal-to-noise ratio, may provide a better correlation with the RV than the chromospheric proxies whose variations tend to level off at very low levels of activity. Moreover, the distribution of the ARs on the photosphere may sometimes differ from that in the chromosphere. In addition to CCF asymmetry indicators, the full width at half maximum (FWHM) of the CCF itself has provided a useful proxy for the activity-induced RV variations in slowly rotating stars \citep[e.g.][]{Dumusque14,Santosetal14,Fischeretal16}, so we shall include it in our investigation. 
We stress that a measurement of the mean line profile distortions by means of the CCF variations would be the ideal activity indicator. It is the CCF  variations that create the measured shift in RV, therefore by characterizing the former we can characterize the latter. This is completely different than associating the RV signal with chromospheric indicators or photometry, which are proxies of activity (i.e. created by the same physical cause that creates RV variations, that is surface magnetic fields) but which, being proxies, might not always be present or correlated with the RV variations.

We analyse a  dataset of spectra of G-type main-sequence stars with known exoplanets to further explore the advantages and the drawbacks of CCF line profile indicators. Our sample consists of stars with a low level of activity and slow rotation ($2 \la v\sin i \la 5$ km\,s$^{-1}$), comparable to the Sun, that have been observed with HARPS-N at the Italian 3.58-m {\it Telescopio Nazionale Galileo} within the coordinated observational programme Global Architecture of Planetary Systems \citep[GAPS; ][]{Covinoetal13,Desideraetal13}. The aim of this specific project is  the detection of additional planets in those systems. While our new discoveries have been reported in other works \citep[e.g.][]{Desideraetal14,Damassoetal15}, here we focus on the study of the correlations of different line profile asymmetry indicators and the FWHM of the CCF with the residual RV variations, that is, those obtained by subtracting the modulations due to the known planets, and whose origin can be confidently attributed to stellar activity.  

\section{Observations}
\label{observations}

Our sample consists of 15 late-type stars hosting planets listed in Table~\ref{table_star_param} together with their effective temperature $T_{\rm eff}$, surface gravity $\log g$, metallicity [Fe/H], projected rotational velocity $v \sin i$,  mean chromospheric Ca~II~H\&K index $\log R^{\prime}_{\rm HK}$ with its standard deviation,  number $N_{\rm RV}$ of RV measurements considered for the present investigations (see below), and  their total time span $\Delta t$ (see below). {The values of $T_{\rm eff}$, $\log g,$ and [Fe/H] were extracted from the reference in the last column, while the values of the $v \sin i$ came from the exoplanets.org database~\citep{Hanetal14}, except for XO-2S for which we used the parameters obtained by  \citet{Damassoetal15b} and \citet{Biazzoetal15}, and HD~108874 for which we used \citet{Benattietal17}.  Typical uncertainties in $\log g$ range between 0.06 and 0.12, while for [Fe/H] they range between 0.03 and 0.07. We used the spectra gathered within the GAPS programme to compute the mean value of $\log R^{\prime}_{\rm HK}$ and its standard deviation  following the method described by \citet{Lovisetal11}. }

HARPS-N is a fiber-fed cross-disperser echelle spectrograph with a fixed setup covering the spectral range 383-690 nm with a resolution of 115\,000  \citep{Cosentinoetal12,Cosentinoetal14}.  We used exposure times of 600 or 900~s yielding a signal-to-noise ratio (hereafter S/N) between $\sim 80$ and $\sim 210$ at 550~nm, according to the apparent brightness  of the target. The data reduction software (hereafter DRS) of HARPS-N is the pipeline used to obtain the RV by cross-correlating the observed spectrum with a binary weighted line mask.
These masks are chosen from templates matching the spectral types of the observed stars and the method used to compute the CCF is described in \citet{Baranneetal96}.
The DRS provides also the CCF profile, which is used to compute the asymmetry indicators and the FWHM as explained in Sect.~\ref{methods}.  The median error of the RVs as returned by the DRS ranges from 0.45 to 1.06~m\,s$^{-1}$ for the stars in our sample, except for X02S for which it is 2.34~m\,s$^{-1}$. 

HARPS-N is enclosed in a controlled environment that ensures a stability of the RV measurements better than 1 m\,s$^{-1}$ along several years. The uniform illumination of the entrance pupil and the stability of the system provide also a very stable CCF that can be used for our purposes. A very small variation of the focus of the instrument was detected during the first period of its operation. This variation perturbed the CCF by slightly modifying its FWHM without appreciably affecting the RV measurements. The focus was stabilized and the FWHM trend eliminated starting from JD~2456738. Therefore, we consider only spectra taken after that date with a time baseline of about two years. We discard a few measurements obtained at large airmass ($\ga 1.8$) and/or at small signal-to-noise ratio (S/N~$\la 25-30$ at 550 nm).

We subtracted the orbital motion of the known planets from the RV time series of each star, by performing an independent fit of the joined literature and HARPS-N datasets. 
The number of planets in each system was taken from the literature with the addition of an independent RV offset and a jitter for each dataset, and of a linear long-term trend.
We computed the RV residuals by choosing the median of the posterior distributions of the orbital parameters, computed with the \textrm{PyORBIT}\footnote{Available at \url{https://github.com/LucaMalavolta/PyORBIT}.} code as described in 
\citet{Malavoltaetal16}. The analysis of the full RV time series and the corresponding implications in terms of orbital parameters of known planets and detectability of additional companions will be presented in a forthcoming study.  For the stars in our sample, the RV modulations induced by the known planets are at least an order of magnitude greater than the RV fluctuations produced by activity, thus the impact of the latter on the orbital parameters of the known planets is small. Nevertheless, a bias is introduced in the RV residuals when we fit a  keplerian model because the Fourier components of the stellar activity variation at the orbital periods of the planets and their harmonics can affect the keplerian fit. 
To quantify this effect, we repeated the analysis by including in the model a linear correlation between the RV variation and the index of chromospheric activity $\log R^{\prime}_{\rm HK}$, at least for the subsets for which this activity indicator is available, and compared the results with the keplerian model without this correlation (see Sect.~\ref{results}).

The residuals of the model including the linear correlation between the RV and the chromospheric index were searched for periodicities longer than 1.0~d arising from possible additional planets  by means of the Generalized Lomb-Scargle periodogram of \citet{Zechmeisteretal09}. No significant periodicities with a false-alarm probability (FAP) $< 0.001$ were found. The  FAP estimated with the analytical formula of \citet{Zechmeisteretal09} was checked by computing the periodograms for $10\,000$ random shufflings of the RV residuals, while the times of the observations were held fixed.   Therefore, we assume that the residual RV time series are dominated by the intrinsic RV variation of the stars and use them to study their correlations with the activity indicators as explained in Sect.~\ref{methods}. 

The standard deviations of the RV residuals of the models including only the keplerian motions and a long-term linear trend in the RV range from $\sim 1.47$~m\,s$^{-1}$ for HD~99109 to $\sim 5.73$~m\,s$^{-1}$ for HD~75898, the latter being one of our fastest rotators with $v \sin i = 4.5 \pm 0.5$~km~s$^{-1}$ (cf. Table~\ref{table_star_param}). {Considering the models including also the correlation between the RV and the activity index, we see 
that this correction practically affects only the residuals of HD~75898 because their standard deviation is decreased to $\sim 5.58$~m\,s$^{-1}$. In the case of the Sun, the standard deviation of the RV variations ranges between $2.46$ and $3.44$~m\,s$^{-1}$ on comparable timescales, depending on the  level of activity \citep[cf. Sect.~4.2 of][]{Lanzaetal16}, which is similar to the average of the present sample. 
\begin{table*} 
\caption{Sample of stars for which we consider asymmetry indicators and FWHM of the CCF. }
\begin{center}
\begin{tabular}{lcccccccc}
\hline
Name & $T_{\rm eff}$ & $\log g$ & [Fe/H] & $v \sin i $ & $\log R^{\prime}_{\rm HK}$ & $N_{\rm RV}$ & $\Delta t$~ & Reference \\
   & (K) & (cm~s$^{-2}$) & & (km\,s$^{-1})$ & & &  (days) & \\
\hline
 & & & & & & & &  \\
 \object{HD~11506} & $6204 \pm 50$ & 4.44 & +0.36 & $5.0 \pm 0.5$  & $-5.028 \pm 0.017$ & 37 & 678.32 & \citet{Santosetal13}  \\
 \object{HD~13931} & $5830 \pm 45$ & 4.30 & +0.03 & $2.0 \pm 0.5$  & $-5.043\pm 0.012$ & 67 & 708.34 & \citet{ValentiFischer05} \\
 \object{HD~23596} & $6110 \pm 45 $ & 4.25 & +0.31 & $4.2 \pm 0.5 $ & $ -5.039 \pm 0.020$ & 26 & 737.36  & \citet{Santosetal13} \\
 \object{HD~72659} & $5920 \pm 45$ &4.24 & 0.0 & $2.2 \pm 0.5$ & $-4.998 \pm 0.019$ & 23 & 737.35 & \citet{ValentiFischer05} \\
 \object{HD~73534} & $5040 \pm 65$ & 3.78 & +0.23 & $ 0.5 \pm 0.5$ & $-5.242 \pm 0.028$ & 31 & 737.38 & \citet{Valentietal09} \\
 \object{HD~75898} & $6140 \pm 50$ & 4.31 & +0.30 & $4.5 \pm 0.5$ & $-5.024 \pm 0.032$ & 34 & 737.47 & \citet{Santosetal13}\\
 \object{HD~89307} & $5900 \pm 45$  & 4.34 & $-0.16$ & $3.2 \pm 0.5$ & $-4.970 \pm 0.008$ & 19 & 737.46 & \citet{ValentiFischer05}\\ 
 \object{HD~99109} & $5270 \pm 45$ & 4.44 & +0.31 & $1.9 \pm 0.5$ & $-5.139 \pm 0.037$ & 27 & 754.38 & \citet{ValentiFischer05} \\
 \object{HD~106252} & $5870 \pm 45$ & 4.36 & $-0.08$ & $1.9 \pm 0.5$ & $-5.003 \pm 0.008$ & 44 & 754.42 & \citet{ValentiFischer05}\\
 \object{HD~108874} & $5585 \pm 20$ & 4.39 & +0.19 & $1.4 \pm 0.3$ & $-5.050 \pm 0.027$ & 55 & 754.46 & \citet{Benattietal17}\\
 \object{HD~155358} & $5900 \pm 100$ & 4.16 & $-0.51$ &  $\sim 2$   &  $-4.965 \pm 0.009$ & 44 & 753.54 & \citet{Robertsonetal12}\\
 \object{HD~188015} & $5745 \pm 45$ & 4.44 & +0.29 & $\leq 0.5$ & $-4.988 \pm 0.027$ & 22 & 737.76 & \citet{ValentiFischer05} \\
\object{HD~190228} & $5350 \pm 45$ & 3.98 & $-0.18$ & $1.9 \pm 0.5 $ & $-5.125 \pm 0.012$ & 44 & 553.42 & \citet{ValentiFischer05}\\
\object{HD~220773} & $5940 \pm 100$ & 4.24 & +0.09 & $\sim 3$ & $-5.088 \pm 0.010$ & 46 & 653.30 & \citet{Robertsonetal12} \\
\object{XO-2S} & $5395 \pm 60 $ & 4.43 & +0.39 & $1.7 \pm 0.4$ & $-5.053 \pm 0.075$ & 44 & 737.39 & \citet{Damassoetal15}\\
  & & & & & & & & \\
  \hline
  \label{table_star_param}
 \end{tabular}
\end{center}
\end{table*}

\section{Methods}
\label{methods}

Several indicators of spectral line asymmetry have been proposed in the literature \citep[e.g.][]{Boisseetal11,Figueiraetal13}. The bisector inverse span (hereafter BIS) is probably the most widely used and proved to be well correlated with the activity-induced RV variations of active stars with $v \sin i \ga 7-10$ km\,s$^{-1}$ \citep[e.g.][]{Quelozetal01}. In stars with a smaller rotation velocity and a lower level of activity, the BIS is not always the best asymmetry indicator and other indicators have been explored \citep{Figueiraetal13}. Following those  results, in addition to the BIS, we consider the two CCF asymmetry indicators $\Delta V$ of \citet{Nardettoetal06} and $V_{\rm asy}$ of \citet{Figueiraetal13}, the latter designed to fully exploit the information content of the different portions of the CCF (see Sect.~\ref{indic_def_err} for their definition). 

All the stars in our sample have $ v\sin i \la 5$ km\,s$^{-1}$ and a typical chromospheric index $ \log R^{\prime}_{\rm HK} \la -4.95$, that is, an activity level comparable with that of the Sun {close to the minimum} of the eleven-year cycle.\footnote{According to  \citet{Dumusqueetal11sun}, $\log R^{\prime}_{\rm HK} $ ranges from $-5.0$ at minimum to $-4.75$ at maximum along  the activity cycle of the Sun.} {The small filling factor of their active regions ($\la 1$ percent) and the small rotational broadening of their line profiles in comparison to the width of the spectrograph instrumental profile make line distortions barely detectable, thus making all asymmetry indicators remarkably less sensitive than in more rapidly rotating and more active stars}. In that case, the FWHM may become a more sensitive activity proxy, so we consider it in addition to the line asymmetry indicators.  The better performance of the FWHM is likely related to the dominance of facular areas in low-activity stars \citep[e.g.][]{Dumusque14,Dumusqueetal14}. Faculae induce a variation of the convective shifts of the line profiles leading to a perturbation that is not localized in the core of the lines, but affects to some extent also their wings and therefore the FWHM of the CCF. } 

\subsection{Asymmetry indicators and FWHM of the CCF}
\label{indic_def_err}

To compute the  BIS, we use the CCF profile provided by the DRS and expressed as the {cross-correlation value} per RV bin of width 250~m$\,$s$^{-1}$. 
We fit a Gaussian profile with a variable continuum level to the CCF to determine the continuum level itself and use it to normalize the CCF profile. The minimum of the normalized CCF profile gives the depth $D$ of the CCF, {also referred to as the CCF contrast}. It is divided into 100 intervals and, for the flux level corresponding to each interval, the RV abscissa is {quadratically} interpolated on the red  and the blue wings of the profile, respectively. The bisector at a given flux level is defined as the arithmetic mean of the RV corresponding to the flux level on the red and blue wings, respectively. Finally, the BIS is computed as the difference of  the mean RV of the bisectors in the top and the lower parts of the CCF, where the top part consists of the elements between 10 and 40 percent of $D$ below the continuum, while the lower part is taken between 60 and 90 percent of $D$  below the continuum. 

In principle, the error of the BIS can be computed by considering the maximum of two different estimates, that is the mean standard error of  the 30  differences between the bisector in the top and the lower parts of the CCF and $\sqrt{2}$ times the formal RV error as given by the DRS.  This approach assumes that: a) all the 30 differences between the top and lower parts of the bisector are independent, which is not the case because the points along the CCF are correlated owing to the cross-correlation procedure giving the CCF itself; b) $\sqrt{2}$ times the formal RV error is a correct estimate of the error in an ideal case, that is, for a CCF without  correlations among its points. Therefore, it is better to assume a more conservative estimate of the error. We increase the error on the bisector differences by a factor of $\sqrt{3}$,  which corresponds to the assumption that there are only ten uncorrelated values along the CCF intervals used to estimate the BIS, and increase from $\sqrt{2}$ to $2.5$ the multiplicative factor applied to the RV error as given by the DRS. {This factor is slightly larger than the generally adopted factor of 2.0 \citep[cf. Sect.~4 of][]{Boisseetal09}, yielding a conservative estimate of the BIS uncertainty.} The maximum of those two estimates is assumed as our error on the BIS. An exact estimate of the number of independent bins along the CCF profile is difficult to obtain, so we follow \citet{Figueiraetal15} who estimate an oversampling factor of $\sim 3$  on the average (see below). In Sect.~\ref{results} we shall compare the errors on the BIS computed with the above procedure with the residuals of the regressions of the BIS time series to quantify the level of error overestimation introduced by our assumptions. 

To compute the indicator $\Delta V$ of \citet{Nardettoetal06} according to the definition in Sect.~5.1 of \citet{Figueiraetal13}, a Levenberg-Marquardt minimization algorithm is applied to fit a bi-Gaussian and a Gaussian profiles to the normalized CCF, respectively. Only the portion of the CCF below 0.95 of the continuum is fitted {to reduce the contribution of  the Lorentzian wings of the photospheric lines to the best fit}. The bi-Gaussian fit has four free parameters: depth, central RV,  FWHM,  and an asymmetry measure.  The Interactive Data Language (hereafter IDL) procedure mpfit.pro\footnote{\url{http://purl.com/net/mpfit}.} is used to compute the best fits.
The indicator $\Delta V$ is defined as  the difference in the central RV values obtained with the Gaussian and the bi-Gaussian best fits, respectively. 
 {With the Gaussian best fit, we also compute the FWHM of the CCF and check that its value is within 0.1 percent of  that provided by the DRS. 
 
The errors on the parameters of the Gaussian and bi-Gaussian best fits, from which  $\Delta V$ and the FWHM of the CCF are derived, cannot be properly estimated from the covariance matrixes given by mpfit.pro because the shape of the CCF is not a Gaussian, which leads to  systematic residuals when we perform a Gaussian best fit. Those systematic residuals are generally greater than the random and the correlated errors of the CCF and produce an overestimate of the errors on $\Delta V$ and the FWHM by a factor of five to seven when we use the standard method based on the covariance matrixes to evaluate their errors. Therefore, we decided to apply a different approach that takes into account both the effect of the random noise and the correlated noise. 
 
The standard deviation $\sigma_{A_{0}(i)}$ of the CCF in the $i$-th flux bin  can be conservatively assumed to be $\sigma_{A_{0}(i)} = \sqrt{A_{0}(i)}$, where $A_{0}(i)$ is the value of the CCF in the bin. This happens because of the method applied to compute the CCF. Specifically, in the photon-dominated regime, the standard deviation of a given spectral element of the spectrum with $N_{\rm ph}$ photon counts is $\sqrt{N_{\rm ph}}$. When we compute the CCF value in a given bin, we add the photon counts in different spectral elements weighted according to the line mask adopted to compute the cross-correlation. Since the mask weights are smaller than the unity,  $A_{0}(i)$ is smaller than the sum of the photoelectron counts in the spectral elements contributing to that bin of the CCF, implying that the signal-to-noise ratio of the CCF is reduced with respect to that corresponding to the total photoelectron counts. Therefore, our prescription to compute $\sigma_{A_{0}(i)}$ is conservative.
  
  To evaluate the effect of the correlated noise along the CCF, first we compute a smoothed version of the CCF itself,  using a Savitzy-Golay filter of order four with ten points on the left and ten on the right of each point as explained in Ch.~14.9 of \citet{Pressetal02}. The residuals between the CCF and its smoothed version are used to compute 100 realizations of the CCF with correlated noise by means of the prayer-bead method \citep[see Sect.~4.2 in][]{Cowanetal12}. Computing $\Delta V$ and the FWHM for all of these 100 realizations, we can evaluate their standard deviations  as produced by the correlated noise. In addition, we compute another 100 realizations by summing to the normalized CCF one hundred realizations of a Gaussian random noise of zero mean and standard deviation $1/ \sqrt{A_{0}(i)}$. Finally, we sum in quadrature the standard deviations of $\Delta V $ and the FWHM for the 100 realizations with correlated noise and the 100 realizations with random noise, thus obtaining their standard deviations that include the effects of both the correlated and the random noises. }

The $V_{\rm asy}$ indicator of Figueira et al. (2013) incorporates the radial velocity $RV(i)$ of each CCF element into its definition. Therefore, it changes if the RV scale is shifted by a constant amount or if the star is genuinely Doppler shifted due to the presence of a planet, thus giving rise to a misleading rejection of any true orbital motion of the star. Moreover, the value of $V_{\rm asy}$ is not invariant when the number of photoelectron counts {along the spectrum} varies, as is the case when a different exposure time is adopted or the star is observed at different air masses. These drawbacks were addressed by \citet{Figueiraetal15} who revised the definition of the indicator.  Here we propose an  improved definition  of $V_{\rm asy}$ to eliminate such spurious dependences and make the indicator non-dimensional, that is, 
\begin{equation}
V_{\rm asy (mod)} = \frac{\sum_{i} \left[ W^{\prime}_{i}({\rm red}) -W^{\prime}_{i}({\rm blue}) \right] \times \overline{W^{\prime}}_{i}}{\sum_{i} \overline{W^{\prime}}^{2}_{i}},
\label{eq2}
\end{equation}
where the weight at flux level $i$ on the red or the blue wings of the CCF  is defined as 
\begin{equation}
W^{\prime}_{i} = \frac{1}{A_{0}(i)} \left[ \frac{\partial A_{0}(i) }{\partial RV(i)} \right]^{2}, 
\end{equation}
where $A_{0}(i)$ is the cross-correlation value at flux level $i$  and $\partial A_{0}(i) / \partial RV(i)$ the derivative of the CCF.  The derivative   is computed using a Savitzy-Golay filter of order four with ten points on the left and ten on the right of each given point. 
 The mean weight $\overline{W^{\prime}}_{i}$ for a given flux level is the arithmetic mean of the corresponding weights on the red and blue wings of the CCF. To avoid sampling too close to the continuum or to the bottom of the CCF, the summations in Eq.~(\ref{eq2}) are extended from five percent to 95 percent of the depth $D$ of the CCF. The  square of the weights in the denominator of Eq.~(\ref{eq2}) makes $V_{\rm asy (mod)}$ invariant for a multiplication of the number of the photoelectron counts by a constant factor. 

To compute the error of $V_{\rm asy (mod)}$, we consider both the effects of a Gaussian noise and of a correlated noise following the procedure previously described for $\Delta V$ and the FWHM. The final value of $V_{\rm asy}$ is the median over the 200 realizations of the CCF obtained with random and correlated noises. The square root of the sum of the variances of the 100 realizations with correlated noise and of the 100 with Gaussian noise is adopted as the standard deviation of the indicator, as in the case of $\Delta V$ and the FWHM. {In light of the above considerations, this is a conservative estimate of the error on our indicator}. 

\citet{Figueiraetal15} consider the effect of the correlated noise in the CCF in the evaluation of their new $V_{\rm asy}$ indicator. Our $V_{\rm asy (mod)} $ is the same as their new indicator provided that, in their notation $\sigma_{\rm CCF}^{2} (i) \propto A_{0}(i)$, that is, that the error on the RV as induced by the flux measurement in the bin $i$ of the CCF is proportional to the {value of the cross-correlation} in that bin. They  recommend choosing the flux grid in such a way that it projects onto a grid of bins separated on the average by $\alpha_{\rm corr}$ bins of the original CCF, where $\alpha_{\rm corr} = (\mbox{size of the bin in RV})/(\mbox{step used in the CCF calculation})$ (see their Sect.~2.1 for details). In the case of the HARPS  spectra, $\alpha_{\rm corr} = 3.28$. This implies that we oversample the CCF by the same factor when we compute our $V_{\rm asy (mod)} $ by considering all the flux levels between five and 95 percent of the continuum. We can correct for this effect by modifying our definition of the weight $W_{i}^{\prime}$ at flux level $i$ as 
\begin{equation}
W^{\prime}_{i} = \frac{1}{\alpha_{\rm corr}} \frac{1}{A_{0}(i)} \left[ \frac{\partial A_{0}(i) }{\partial RV(i)} \right]^{2}. 
\end{equation}
Since the factor $\alpha_{\rm corr}$ is constant and our definition of $V_{\rm asy (mod)}$ is invariant for any constant scale factor applied to the {CCF value} $A_{0}(i)$, our value of the indicator is independent of $\alpha_{\rm corr}$. Therefore, we maintain our previous definition of $V_{\rm asy (mod)}$ and do not introduce any correction for the oversampling associated with the correlated nature of the CCF. We provide an IDL procedure to compute the asymmetry indicators and the FWHM of the CCF, which is described in Appendix~\ref{appendix}.

\subsection{Correlations of the activity indicators with the RV variations}
\label{correlation_coefficients}

Our RV residual time series have no significant periodicities, therefore we cannot compare their periodograms with those of the activity indicators to investigate the correlation between RV  and activity as for example in \citet{Santosetal14}, because our periodograms are dominated by random noise. 
Moreover, previous investigations showed that the correlations between the RV variations and our indicators  generally are neither linear nor monotonic; for example, a plot of the RV versus the BIS generally displays an eight-shaped pattern \citep{Boisseetal11,Figueiraetal13}. From this point of view, we are far from the ideal case of linearly correlated or even monotonically correlated indicators, {thus Pearson or Spearman correlation coefficients are of very limited use. 

Recently, Gaussian process (hereafter GP) regression has been proposed to model the RV variations produced by stellar activity \citep[e.g.][]{Haywoodetal14,Rajpauletal15}. It is a non-parametric regression method that allows us to model complex time variations with both stochastic and  deterministic components by parametrizing the covariance between pairs of data points. The covariance is specified by a set of hyperparameters that can be estimated from the dataset within a Bayesian framework \citep[]{Robertsetal12}. 

In the case of RV variations induced by stellar activity, the covariance is specified by a set of four hyperparameters associated with the amplitude of the correlation, the timescale for the growth and decay of active regions, the rotation period of the star, and the smoothing of the rotational modulation. These hyperparameters can be estimated either from RV time series with a suitable cadence and extension \citep[e.g.][]{LopezMoralesetal16} or using external datasets with high cadence, for example high-precision  photometry \citep[e.g.][]{Malavoltaetal18}. 

 In the present work, we explore an alternative approach to model the regression between the RV variations and the activity indicators. 
Specifically, we model the dependence of the RV on time and activity indicators by means of kernel regression (hereinafter KR), another non-parametric regression technique. Its foundations are given in, for example, \citet{Hardle92}, \citet{Loader99}, and \citet{Takedaetal07}, while previous applications to the analysis of astronomical data can be found in \citet{Wangetal07}, \citet{Marcoetal15}, or \citet{ALOtaibietal16}.  

We apply a locally linear model to fit the RV at a time $t_{k}$ by minimizing the objective function $Z$:
\begin{equation}
Z \equiv \sum_{i=1}^{N} \left[ RV(t_{i}) - \beta_{0} - \beta_{1} (x_{i} - x_{k}) \right]^{2} W(x_{i}-x_{k}, t_{i}-t_{k}, h_{x}, h_{t}),
\label{kr_eq}
\end{equation}
where $N$ is the number of RV and simultaneous activity indicator observations, $RV(t_{i})$  the RV at time $t_{i}$, $x_{i} \equiv x(t_{i})$  a generic  indicator at time $t_{i}$, $\beta_{0}$ and $\beta_{1}$ the coefficients with respect to which the objective function $Z$ is minimized, and $W$  the kernel given by
\begin{equation}
W(x_{i}-x_{k}, t_{i}-t_{k}, h_{x}, h_{t}) = \exp \left\{ - \left[ \left( \frac{x_{i}-x_{k}}{h_{x}} \right)^{2} + \left( \frac{t_{i} -t_{k}}{h_{t}} \right)^{2} \right] \right\}, 
\label{kernel_eq}
\end{equation}
where $h_{x}$ and $h_{t}$ are the bandwidths of the kernel. Equation (\ref{kr_eq}) is a linear regression of the RV versus the activity proxy $x$ where the datapoints are weighted according to the kernel in Eq.~(\ref{kernel_eq}). This gives more weight to points closer in time to $t_{k}$ and to the indicator $x(t_{k})$, thus  accounting for the temporal variation of the correlation and the non-linear dependence of the RV on the indicator itself. By the standard method to compute a weighted linear best fit \citep[e.g.][]{Pressetal02}, we find the kernel regression estimator for the radial velocity at the time $t_{k}$ as \begin{equation}
\hat{RV}(t_{k}) = \sum_{i=1}^{N} M_{ki} \, RV(t_{i}),
\end{equation}
where the elements of the matrix ${ \vec M}$ are given by
\begin{eqnarray}
M_{ki} &  = & \frac{S_{2}(x_{k}) W(x_{i}-x_{k}, t_{i}-t_{k}, h_{x}, h_{t})}{\Delta(x_{k})} + \nonumber \\ 
&  & -\frac{S_{1}(x_{k}) (x_{i}-x_{k})W(x_{i}-x_{k}, t_{i}-t_{k}, h_{x}, h_{t})}{\Delta(x_{k})},
 \end{eqnarray} 
 where 
 \begin{equation}
 S_{l} (x) = \sum_{i=1}^{N} W(x_{i}-x, t_{i}-t, h_{x}, h_{t})(x_{i} - x)^{l},
 \end{equation}
with $l = 0, 1, 2$ and
\begin{equation}
\Delta(x) = S_{2} (x) S_{0} (x) - [S_{1}(x)]^{2}.
\end{equation}

The optimal values of the bandwidths $h_{x}$ and $h_{t}$ are obtained by the so-called leave-one-out method \citep{Hardle92}, that is, by minimizing the function
\begin{equation}
C(h_{x}, h_{t}) = \frac{1}{N_{\rm r}} \sum_{i=1}^{N_{\rm r}} \left[ \frac{RV(t_{i}) - \hat{RV}(t_{i})}{1 - M_{ii}} \right]^{2} 
\label{leave-one-out-eq}
\end{equation}
with respect to $h_{x}$ and $h_{t}$, where the summation is made on the $N_{\rm r}$ datapoints that have a distance in $x$ and $t$ of at least $ 2h_{x}$ and $2h_{t}$ from the extreme points of the ranges in $x$ and $t$, respectively, to avoid the effect of the worse quality of the fit at the border of the domain.\footnote{The justification of Eq.~(\ref{leave-one-out-eq}) can be found in, e.g., \citet{Loader99} or Sect.~5.2 of the notes by R.~Tinshirani at \url{http://www.stat.cmu.edu/~larry/=sml/nonpar.pdf}.} We impose that $h_{x}$ and $h_{t}$ be smaller than $1/8$ of the total range of the indicator $x$ and of the time interval covered by the observations, respectively, to have a sufficient resolution of the kernel. Moreover, we impose that $h_{x}$ and $h_{t}$ be  greater than 4 times the mean separation of the datapoints in $x$ and $t$  to avoid a too small number of datapoints effectively contributing to the regression at a given point. To perform the constrained minimization, we make use of the IDL procedure {\tt CONSTRAINED\_MIN}. 

In addition to fitting the RV observations, KR can be used to interpolate the RV values between the times of the actual observations. To this purpose, first we linearly interpolate the values of $x$ on an evenly sampled time grid $\{t_{j}\}$ and then compute the kernel regression over the couples $(t_{j}, x(t_{j}))$  to obtain a plot without time gaps.  Confidence intervals for the estimator $\hat{RV}(t)$ are computed by  the method in Sect.~ 2.3.3 of \citet{Loader99}. 

The significance of the KR with respect to a given indicator $x$ and the time $t$ can be derived by a suitable application of the Fischer-Snedecor $F$ statistics as discussed in Ch.~9 of \citet{Loader99}. Specifically, we compute the ratio-of-variance statistics $F$ to compare the kernel regression model with the model assuming no dependence of the RV on the given indicator as
\begin{equation}
F= \left( \frac{N-\nu-1}{\nu -1} \right) \frac{\sum_{i} [RV(t_{i})-\overline{RV}]^{2}- \sum_{i}[RV(t_{i})-\hat{RV}(t_{i})]^2}{\sum_{i}[RV(t_{i})-\hat{RV}(t_{i})]^{2}},
\end{equation}
where $\overline{RV} = \sum_{i} RV(t_{i})/N $ is the mean of the RV measurements and $\nu$ is the effective number of degrees of freedom of the KR that is given by \citep[cf.][]{Loader99}
\begin{equation}
\nu = {\rm Tr} ({\vec \Lambda}^{\rm T} {\vec \Lambda}), 
\end{equation}
where $\rm Tr$ gives the trace of the argument matrix and ${\vec \Lambda}^{\rm T}$ is the transpose of the matrix ${\vec \Lambda}$ that is defined as ${\vec \Lambda} \equiv {\vec I} - {\vec M}$, with ${\vec I}$ being the identity matrix.  
The statistics $F$  is distributed as the ratio of two $\chi^{2}$ variables with $\nu -1$ degrees of freedom in the numerator and $N-\nu-1$ degrees of freedom in the denominator, respectively. Its significance can be computed, for example,  by means of the {\tt F\_PDF} function of IDL. 

 }

\section{Results}
\label{results}

The results of the application of the KR to the time series of the RV residuals obtained with the model including the keplerian motions of the known planets, a long-term linear trend, and a linear correlation between the RV and the chromospheric index $\log R^{\prime}_{\rm HK}$ are presented in Table~\ref{results_rhkcorr_all} at the end of the paper. A selection including only the cases where the significance of the KR is below $0.01$ is presented in Table~\ref{results_rhkcorr}. We considered only the datapoints that had simultaneous measurements of the RV and of all the five indicators, that is $\log R^{\prime}_{\rm HK}$, BIS, $\Delta V$, $V_{\rm asy(mod)}$, and FWHM, and applied the KR twice, first considering all the datapoints in the time series, then excluding the datapoints that deviated from the regression by more than three standard deviations of the residuals of the first regression. Tables~\ref{results_rhkcorr} and \ref{results_rhkcorr_all} list from left to right, the name of the star, the indicator used together with the time to compute the KR (cf. Eqs.~\ref{kr_eq} and~\ref{kernel_eq}), the number $N_{\rm c}$ of datapoints after the above cross-matching and $3$-$\sigma$ clipping, the standard deviation $\sigma$ of these RV residuals  before the application of the KR, the standard deviation $\sigma_{\rm KR}$  after the subtraction of the KR, the time bandwidth $h_{\rm t}$, the indicator bandwidth $h_{\rm x}$, the effective number $\nu$ of degrees of freedom of the KR, the Fischer-Snedecor function $F$ that compares the regression model with that assuming no correlation between the RV residuals and the indicator and time, and the significance $\alpha$ of the correlation, that is, the probability that a value of $F$ as large as that observed can arise from random statistical fluctuations  (cf. Sect.~\ref{correlation_coefficients}). 
\begin{table*}
\begin{center}
\caption{Results of the KR of the RV residuals of the keplerian model including a linear correlation between the RV and the stellar chromospheric index $\log R^{\prime}_{\rm HK}$ for our stars. Only the cases with significant correlations, that is $\alpha < 0.01$, are listed (see the text).}
\begin{tabular}{lcccccccccc}
\hline
                                    Star name & Indicator  & $N_{\rm c}$ & $\sigma$ & $\sigma_{\rm KR\, res}$ & $h_{\rm t}$ & $h_{\rm x}$ & $\nu$ & $F$ & $\alpha$ \\
                                                      & & & (m/s) & (m/s) & (d) & & & & \\
\hline
     HD23596 &                          BIS  &    25 &        2.642 &        1.445 &       85.161 &    1.655e-03 &        8.879 &        4.500 &     0.005952 \\
     HD23596 &                   $\Delta V$  &    25 &        2.642 &        1.347 &       85.161 &    1.931e-03 &       10.182 &        4.269 &     0.007890 \\
     HD23596 &                         FWHM  &    25 &        2.642 &        1.411 &       85.161 &    1.815e-03 &        8.284 &        5.401 &     0.002504 \\
     HD89307 &                          BIS  &    19 &        3.775 &        1.467 &       84.339 &    1.777e-03 &        8.642 &        6.862 &     0.004178 \\
     HD89307 &          $V_{\rm asy (mod)}$  &    19 &        3.775 &        1.187 &       84.339 &    1.815e-03 &        8.046 &       12.648 &     0.000309 \\
    HD106252 &                          BIS  &    43 &        2.350 &        1.259 &       79.651 &    8.857e-04 &       19.965 &        2.884 &     0.009355 \\
    HD155358 &   $\log R^{\prime}_{\rm HK}$  &    43 &        1.947 &        1.089 &       67.353 &    2.200e-03 &       17.987 &        3.095 &     0.005702 \\
    HD188015 &                         FWHM  &    21 &        3.945 &        1.885 &      110.264 &    3.183e-03 &        8.384 &        5.300 &     0.006108 \\
    HD190228 &          $V_{\rm asy (mod)}$  &    43 &        2.621 &        1.437 &       41.111 &    8.937e-04 &       17.415 &        3.482 &     0.002628 \\
        XO2S &                          BIS  &    42 &        3.534 &        1.513 &       70.129 &    3.774e-03 &       19.207 &        5.312 &     0.000171 \\
\hline
\label{results_rhkcorr}
\end{tabular}
\end{center}
\end{table*}
\begin{table*}
\begin{center}
\caption{Results of the KR of the RV residuals of the keplerian model that does not include the linear correlation between the RV and the stellar chromospheric index for our stars. Only the cases with significant correlations, that is $\alpha < 0.01$, are listed (see the text).}
\begin{tabular}{lcccccccccc}
\hline
                                    Star name & Indicator  & $N_{\rm c}$ & $\sigma$ & $\sigma_{\rm KR\, res}$ & $h_{\rm t}$ & $h_{\rm x}$ & $\nu$ & $F$ & $\alpha$ \\
                                                      & & & (m/s) & (m/s) & (d) & & & & \\
\hline
      HD23596 &                          BIS  &    25 &        3.384 &        1.636 &       85.161 &    1.655e-03 &        8.879 &        6.289 &     0.001154 \\
     HD23596 &                   $\Delta V$  &    25 &        3.384 &        1.606 &       85.161 &    1.931e-03 &       10.182 &        5.176 &     0.003308 \\
     HD23596 &          $V_{\rm asy (mod)}$  &    25 &        3.384 &        1.556 &       85.161 &    2.459e-03 &       11.679 &        4.303 &     0.008842 \\
     HD23596 &                         FWHM  &    25 &        3.384 &        1.665 &       85.161 &    1.815e-03 &        8.284 &        6.748 &     0.000786 \\
     HD89307 &                          BIS  &    19 &        3.765 &        1.446 &       84.339 &    1.777e-03 &        8.642 &        7.064 &     0.003759 \\
     HD89307 &          $V_{\rm asy (mod)}$  &    19 &        3.765 &        1.187 &       84.339 &    1.815e-03 &        8.046 &       12.582 &     0.000316 \\
    HD106252 &                          BIS  &    43 &        2.365 &        1.269 &       79.651 &    8.857e-04 &       19.965 &        2.871 &     0.009616 \\
    HD188015 &   $\log R^{\prime}_{\rm HK}$  &    21 &        4.409 &        1.724 &      110.264 &    1.295e-02 &        9.138 &        7.392 &     0.001748 \\
    HD188015 &                         FWHM  &    21 &        4.409 &        1.902 &      110.264 &    3.183e-03 &        8.384 &        6.868 &     0.002084 \\
    HD190228 &          $V_{\rm asy (mod)}$  &    43 &        2.619 &        1.435 &       41.111 &    8.937e-04 &       17.415 &        3.486 &     0.002610 \\
        XO2S &                          BIS  &    42 &        3.538 &        1.537 &       72.829 &    3.774e-03 &       19.499 &        4.969 &     0.000292 \\

 \hline
\label{results_no_rhkcorr}
\end{tabular}
\end{center}
\end{table*}

The results of the KR as applied to the RV residuals of the model including only the keplerian motions and a long-term linear trend in the RV are listed in Table~\ref{results_no_rhkcorr_all} at the end of the paper, which lists the same quantities as in Table~\ref{results_rhkcorr}. A selection of the results with $\alpha < 0.01$ are reported in Table~\ref{results_no_rhkcorr}. 

The standard deviations of the residuals of the orbital best fits  including or excluding the correlation between the RV and the chromospheric index are almost the same, except for HD~23596, HD~188015, and, to a lesser extent, HD~75898, for which they are smaller for the former model because of the additional free parameters. This confirms that the bias introduced by neglecting the dependence of the RV on activity is generally small for the stars in our sample.\footnote{Our stars are characterized by low levels of activity and rather large orbital RV modulations by their known planets, with orbital periods far from the typical rotation periods of sun-like stars. Specifically, the radial velocity semi-amplitude $K$ ranges between $\sim 15$ and $\sim 140$~m s$^{-1}$, while the orbital periods are between $\sim 120$ and $\sim 4200$ days. The activity-induced RV variations have most of their power at the rotation frequency and its first two or three harmonics \citep[e.g.][]{Boisseetal11}, which is  far from the orbital frequencies and their harmonics for almost all of the known planets orbiting our stars. Therefore, the component of  the activity-induced RV variation that can be absorbed by the orbital fit is small in most of our cases.} However, including the $RV-\log R^{\prime}_{\rm HK}$ linear correlation makes a significant difference in those three cases.  Stars  HD~23596 and HD~188105 show a significant correlation between the RV variation and the activity indicators ($\alpha < 0.01$; see Tables~\ref{results_rhkcorr} and~\ref{results_no_rhkcorr}) suggesting the importance of including such a correlation in the keplerian fits. Thus we mainly refer  to models including the $RV-\log R^{\prime}_{\rm HK}$ correlation in the presentation of our results. 

Considering that we test five different correlations for 15 stars, that is a total of 75 possible correlations, we fix the significance threshold at $\alpha = 0.01$ to have less than one expected spurious correlation arising from statistical fluctuations in our samples. Including the $RV-\log R^{\prime}_{\rm HK}$ correlation in the keplerian model gives seven stars out of 15 with $\alpha < 0.01$, while excluding that correlation gives six stars because the minimum $\alpha$ for HD~155358 increases from $0.0057$ to $0.0247$.  Looking at Table~\ref{results_rhkcorr}, we see that the indicator with the greatest number of significant correlations is the BIS with four cases followed by the new indicator $V_{\rm asy(mod)}$ and the FWHM with two cases, while the chromospheric index $\log R^{\prime}_{\rm HK}$ and  $\Delta V$ have only one. The significance of the KR of HD~75898 with respect to the $V_{\rm asy(mod)}$ and the time is 1.36 percent (see Table~\ref{results_rhkcorr_all}). If we include this further case among the significant correlations, $V_{\rm asy(mod)}$ gives three positive cases. On the other hand, in the case of the keplerian model without the $RV-\log R^{\prime}_{\rm HK}$ correlation, the BIS shows four significant correlations, the $V_{\rm asy(mod)}$ three, the FWHM two, and the $\Delta V$ and $\log R^{\prime}_{\rm HK}$ only one significant correlation (cf. Table~\ref{results_no_rhkcorr}). 

In all the cases, the application of the KR reduces the amplitude of the RV residuals  with their standard deviation ranging between $1$ and $2$~m\,s$^{-1}$, except for HD~75898, our most active star. Therefore,  KR proved to be a suitable regression method to account for the effects of stellar activity on RV variations. In the cases of our significant correlations, the standard deviation of the KR residuals is always lower than $1.5$~m\,s$^{-1}$, except for HD~188015, making a highly significant improvement in the modelling of the RV time series. 

\begin{figure}
\hspace*{-7mm}
 \centering{
 \includegraphics[width=8cm,height=10cm,angle=-90]{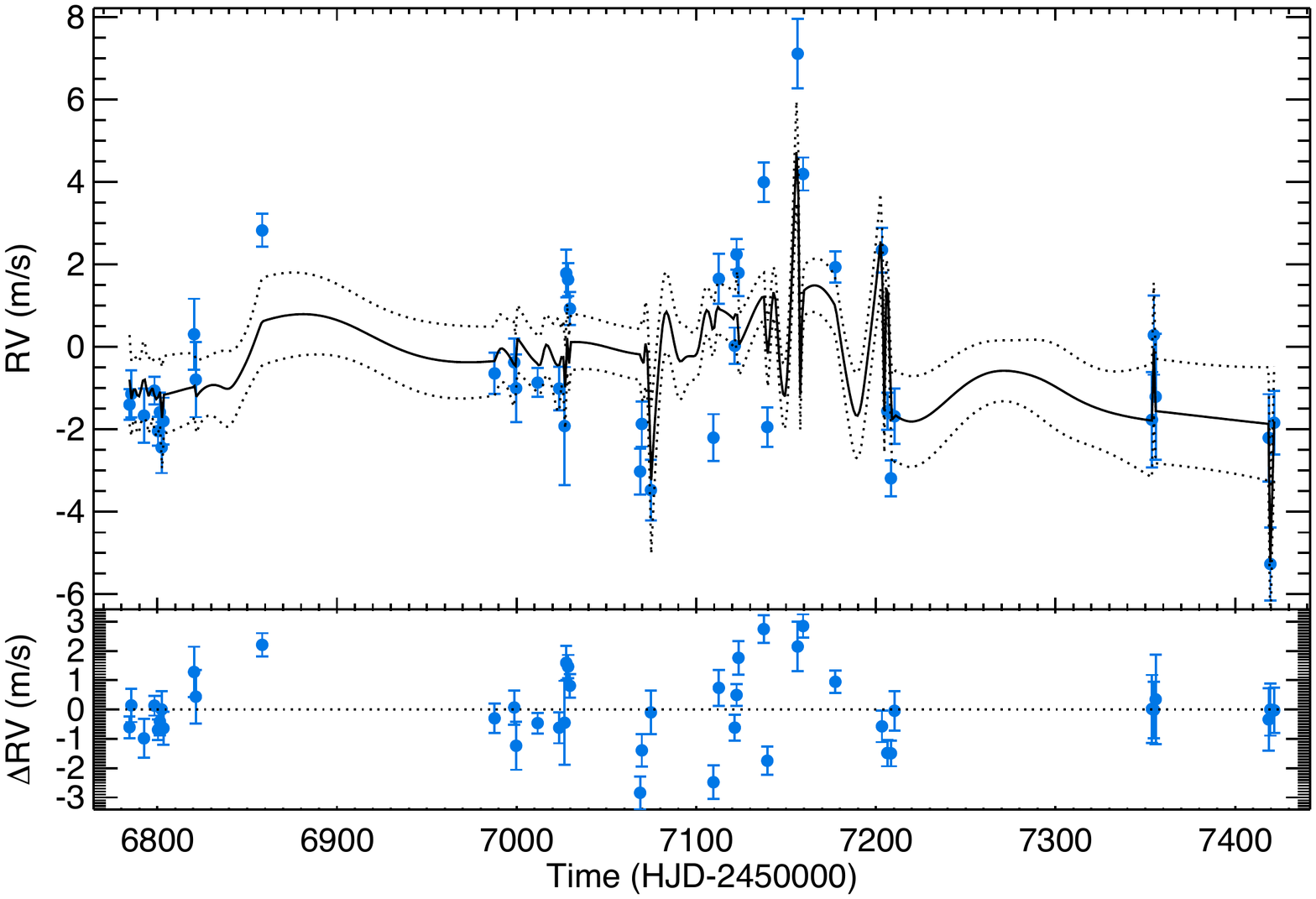}} 
   \caption{Top panel: Radial velocity (RV) residuals of HD~106252 versus time. The solid line is the kernel regression (KR) of the RV time series with respect to the time and the BIS; the $\pm \sigma$ interval of the regression as discussed in Sect.~\ref{correlation_coefficients} is indicated (dotted lines). Bottom panel: RV residuals of the KR plotted in the top panel. Zero residuals are indicated by the dotted line. }
              \label{Fig_hd106252}%
\end{figure}

\begin{figure}
\hspace*{-7mm}
\centering{
 \includegraphics[width=8cm,height=10cm,angle=-90]{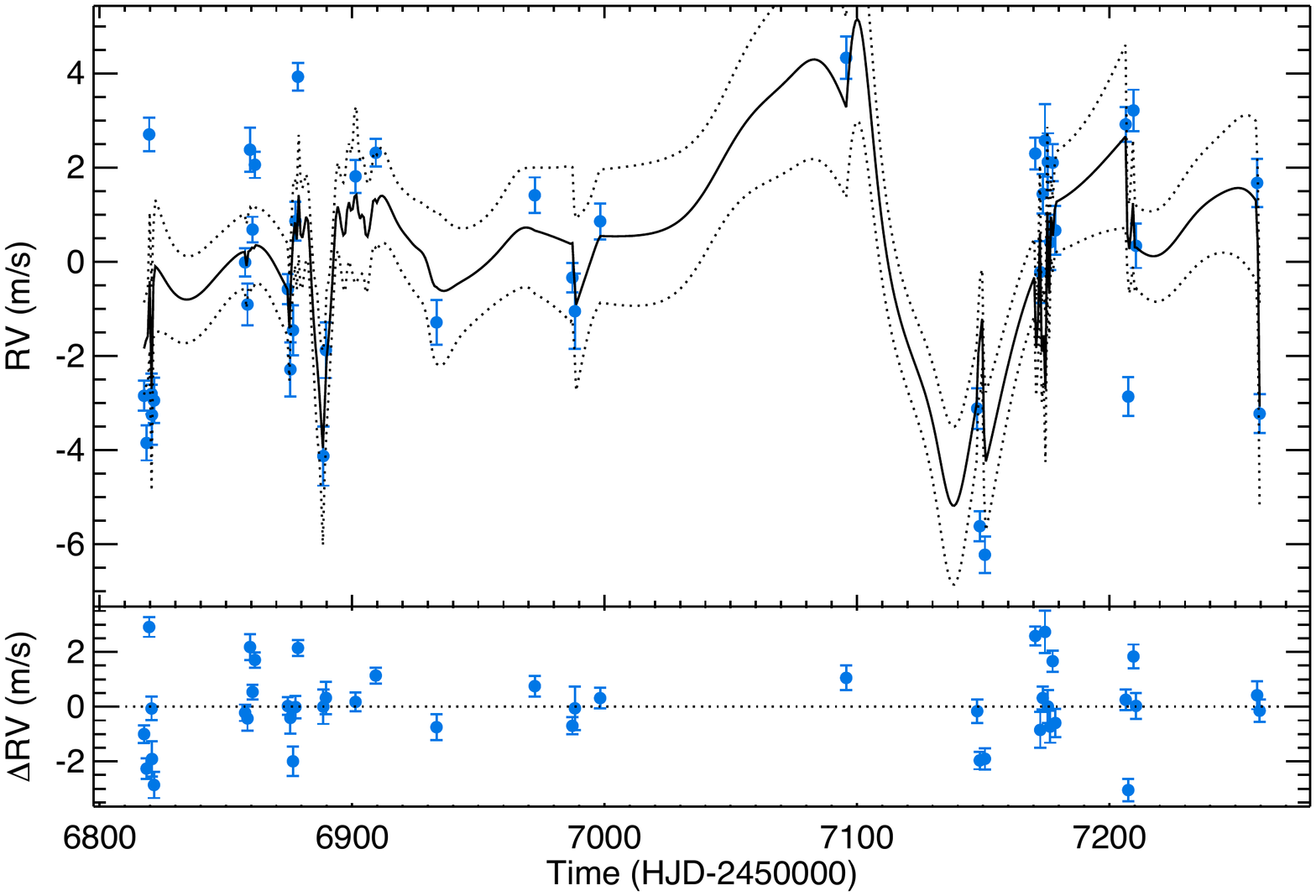}} 
   \caption{Same as Fig.~\ref{Fig_hd106252}, but for HD~190228 with the KR performed with respect to the time and $V_{\rm asy(mod)}$.}
              \label{Fig_hd190228}%
\end{figure}

Two examples of the application of KR to our stars with well sampled time series are plotted in Figs.~\ref{Fig_hd106252} and~\ref{Fig_hd190228} for HD~106252 and HD~190228, respectively, both with 43 datapoints. The slope of the regression changes remarkably, sometimes discontinuously, at times when there are no datapoints to constrain the slope itself or when there are large variations in the RV residuals. Slope changes are more gradual in the time intervals when datapoints are more abundant and show variations with a lower amplitude. This is a consequence of the local linear fit performed on the datapoints and the need to match fits with different slopes. The performance of the regression is generally very good showing that it can capture the non-linear and non-monotonic correlations between the RV variations and the different activity indicators.  We note that  HD~106252 and HD~190228 are targets with  a chromospheric index comparable with that of the Sun at the minimum of the eleven-year cycle.  

\begin{figure}
\hspace*{-7mm}
\centering{
 \includegraphics[width=8cm,height=10cm,angle=-90]{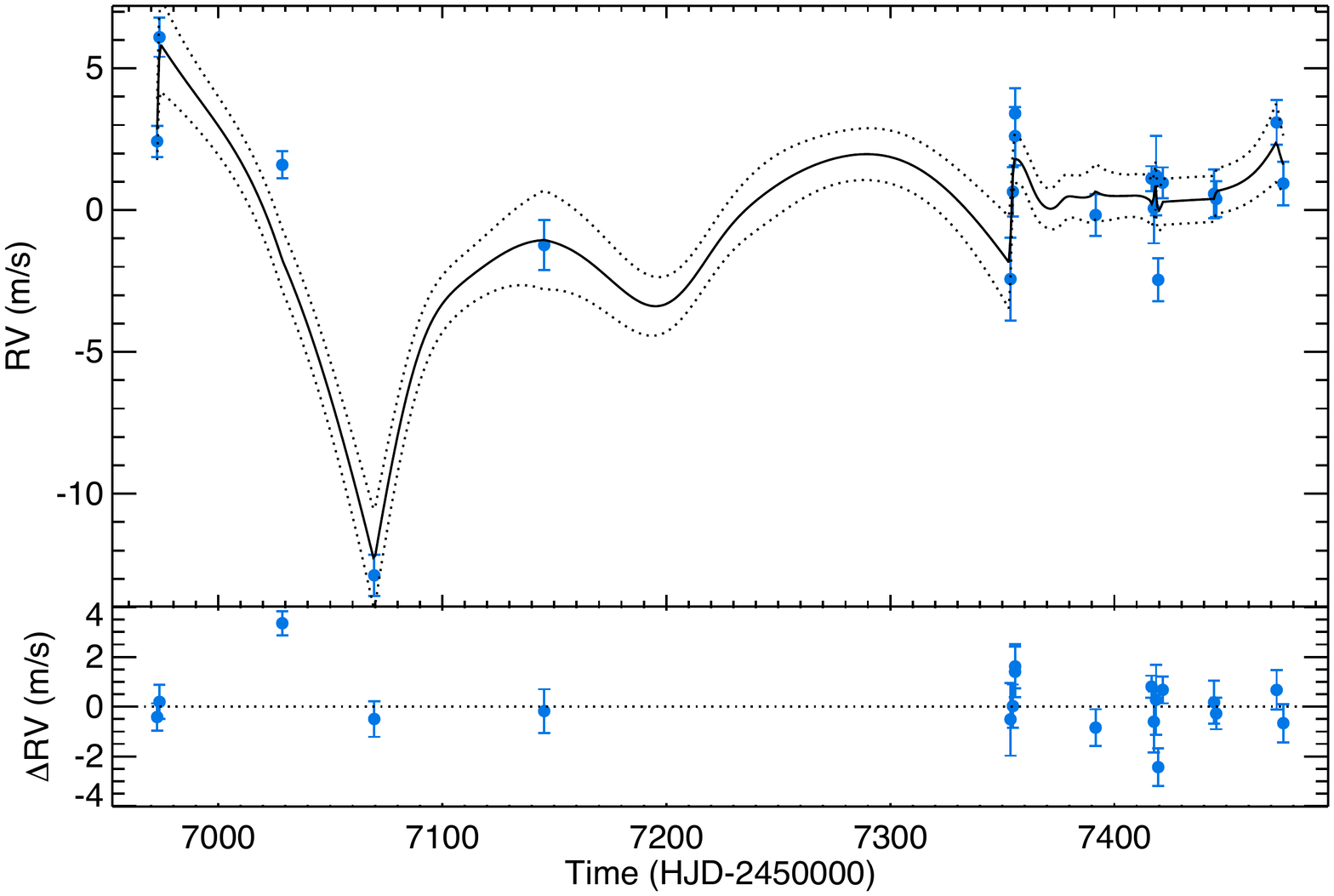}} 
   \caption{Same as Fig.~\ref{Fig_hd106252}, but for HD~89307 with the KR performed with respect to the time and the $V_{\rm asy(mod)}$.}
              \label{Fig_hd89307}%
\end{figure}

\begin{figure}
\hspace*{-7mm}
 \centering{
 \includegraphics[width=8cm,height=10cm,angle=-90]{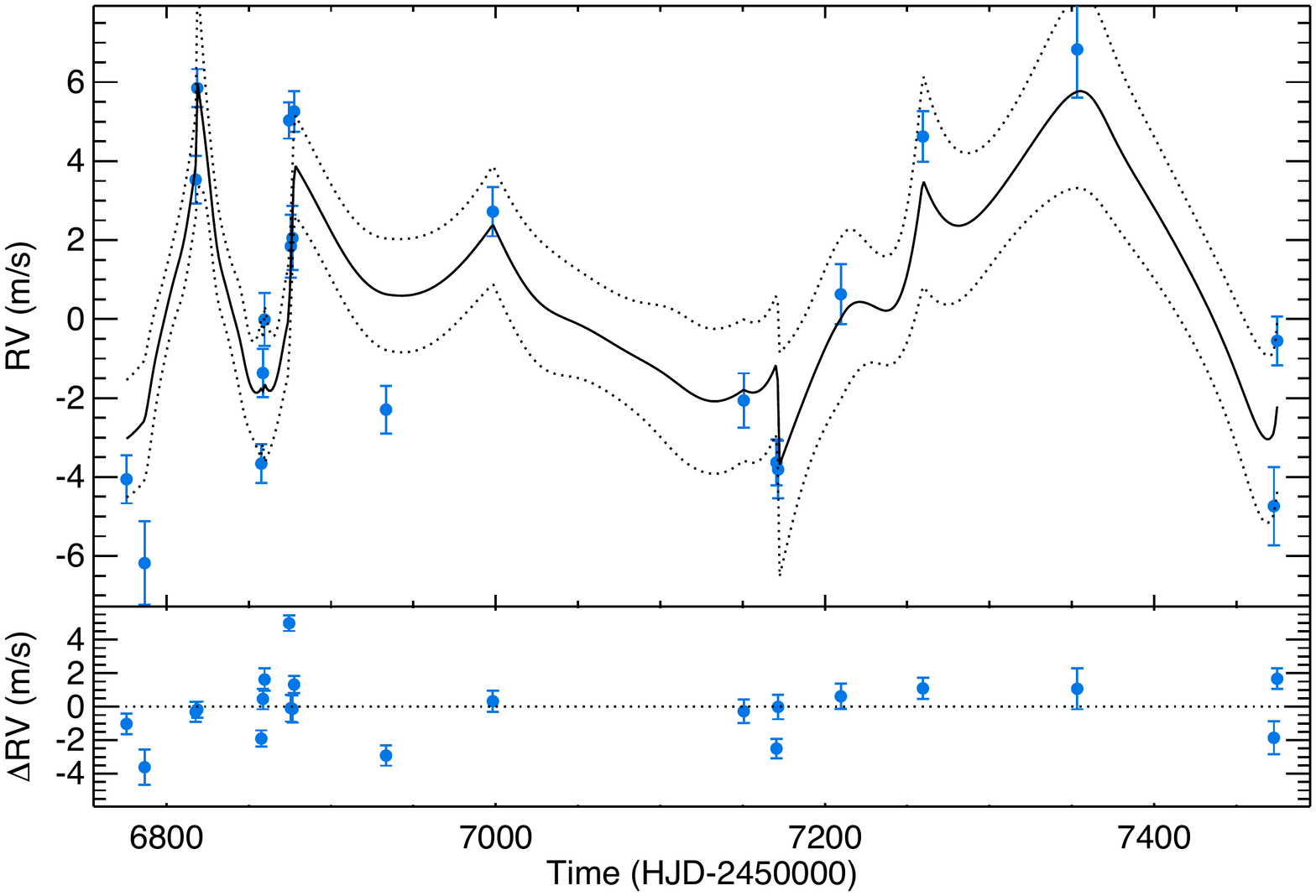}} 
   \caption{Same as Fig.~\ref{Fig_hd106252}, but for HD~188015 with the KR performed with respect to the time and the FWHM.  }
              \label{Fig_hd188015}%
\end{figure}

Two examples of the application of the KR to stars with a relatively low number of datapoints are shown in Figs.~\ref{Fig_hd89307} and~\ref{Fig_hd188015} for HD~89307 with 19 points and HD~188015 with 21 points, respectively.  The number $\nu$ of the degrees of freedom of  the regression model is approximately half the number of the datapoints $N_{\rm c}$ as in the case of the stars with larger datasets indicating that the number of free model parameters $N_{\rm c} -\nu \approx 0.5\, N_{\rm c}$ in both the cases. 


In principle, the bandwidths of our KR can be related to physical properties of the star. For example, $h_{\rm t}$ can be a measure of the evolutionary timescale of the pattern of active regions responsible for the RV variation, or $h_{\rm x}$ can be used to characterize a dependence of the RV perturbation on the level of activity. However, a word of caution is in order here because the sparseness of the sampling of our time series can limit the information on the active region lifetime and the dependence of the RV on the activity level that can be extracted from our datasets. Considering  time series with more datapoints and less gaps such as those of HD~106252 or HD~190228, we estimate an average lifetime of the active regions from $h_{\rm t}$ between $\sim 40$ and $\sim 80$ days, which compares well with the typical lifetimes of solar faculae that dominate the RV variation in the case of the Sun \citep[cf.][]{Meunieretal10,Lanzaetal16,Haywoodetal16}. 

\begin{figure}
\vspace*{6.5mm}
\hspace*{1mm}
 \centering{
 \includegraphics[width=7cm,height=9cm,angle=90]{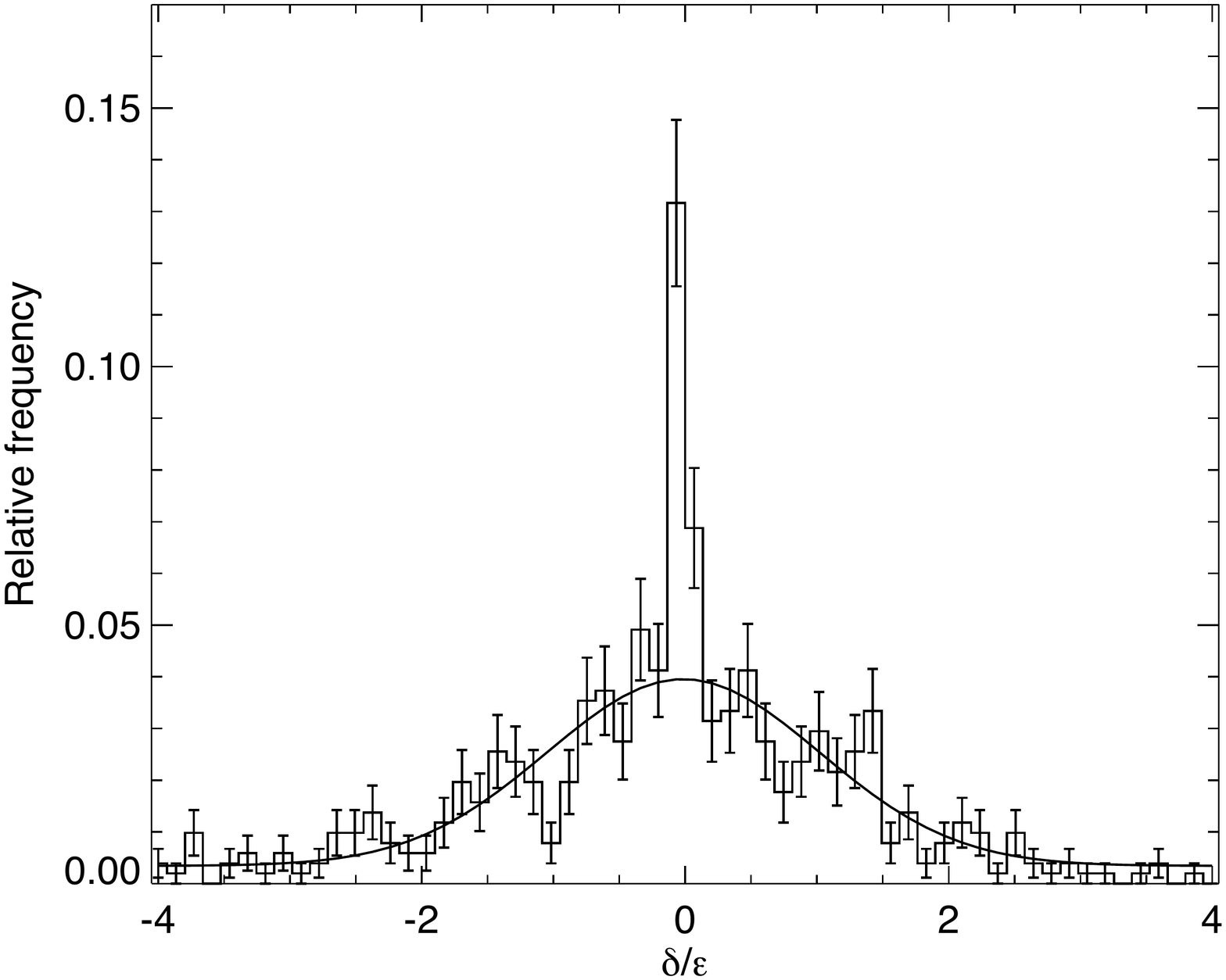}} 
   \caption{Distribution of the ratio $\delta/\varepsilon$ of the KR residuals of  the BIS timeseries of our stars vs. the time and the RV residuals with $\gamma = 0.44$. The solid line is a Gaussian best fit with unity standard deviation computed after excluding the two central bins from the fitting.}%
  \label{BIS_error_KR}
\end{figure}

\begin{figure}
\hspace*{-10mm}
 \centering{
 \includegraphics[width=7cm,height=9cm,angle=90]{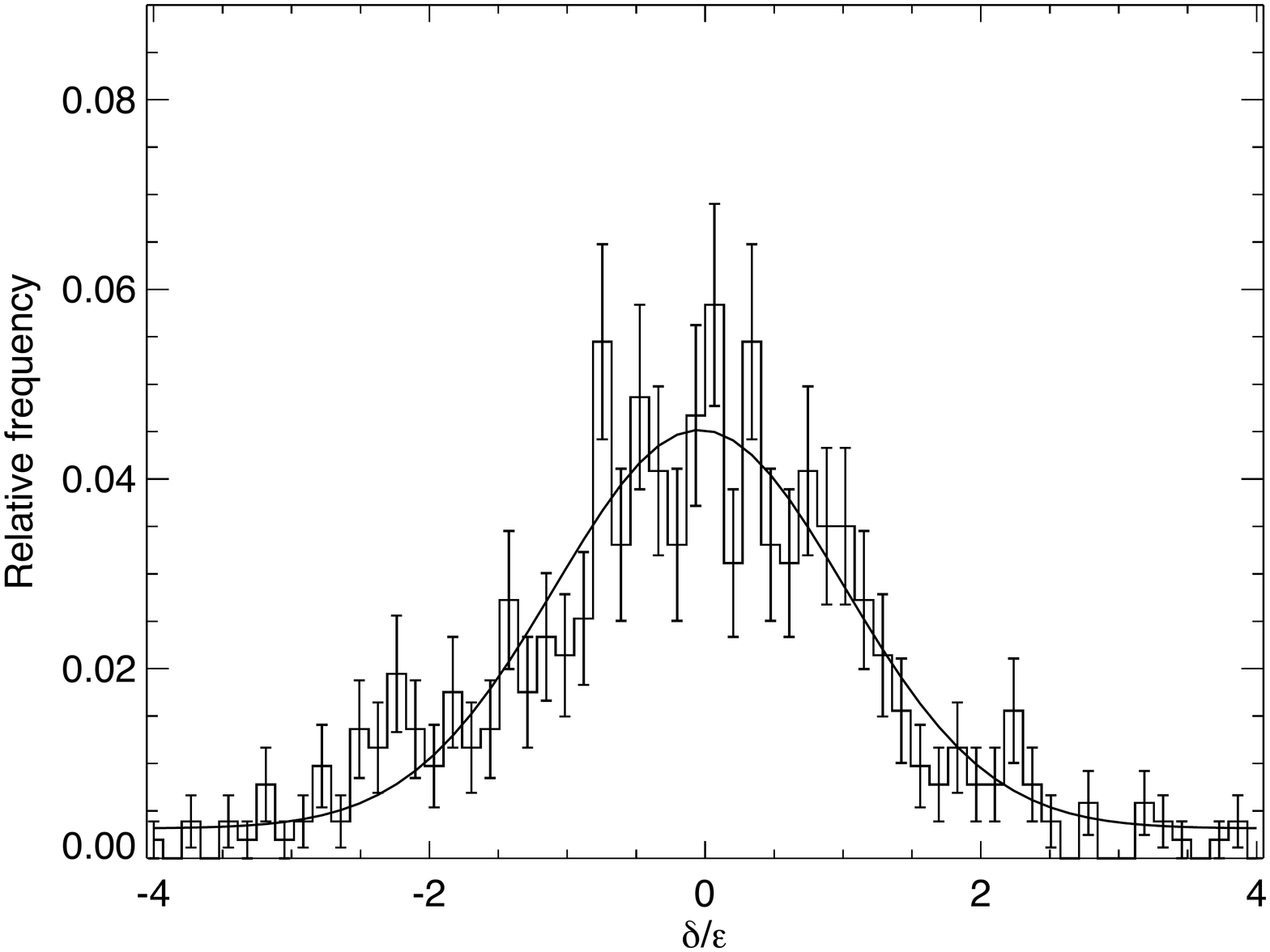}} 
   \caption{Distribution of the ratio $\delta/\varepsilon$ of the residuals of the linear regressions of the BIS timeseries of our stars vs. the time  with $\gamma = 0.66$. The solid line is a Gaussian best fit to the distribution with unity standard deviation. }%
  \label{BIS_error_linear}
\end{figure}

In Sect.~\ref{indic_def_err} we introduced a prescription to estimate the error $\varepsilon$ on the BIS that was designed to overestimate the error itself. We can compare our BIS errors with the residuals of the regressions of the BIS time series of our stars to quantify the degree of overestimation. For each of our stars, we compute the residuals $\delta$ of the KR of the  BIS versus the time and the RV residuals, and consider the distribution of the ratio $\delta/ \varepsilon$ for our whole sample of measurements. Ideally, such a distribution is expected to be a Gaussian with zero mean and unity standard deviation. Actually, we find that we must multiply $\varepsilon$ by a correction factor $\gamma = 0.44$ to obtain a distribution with unity standard deviation as plotted in Fig.~\ref{BIS_error_KR}. We see that the fraction of measurements with very small deviations is significantly greater than expected in the case of a Gaussian distribution. Therefore, the Gaussian best fit plotted in the figure was computed by excluding the two central bins where virtually all the residuals with anomalously small deviations  are concentrated. Those anomalous residuals amount to  approximately ten percent of the total and are a consequence of the overfitting produced by the KR when the number of datapoints is low and their cadence is sparse.  Nevertheless, they do not represent a major problem because their relative fraction is rather small. Alternatively, to avoid any problem with overfitting, we computed a simple linear regression of each of the BIS time series versus the time only. The quality of those regressions is significantly worse than in the case of the KR, but we see no overabundance of residuals in the bins closer to zero. We find that the distribution of the ratio $\delta/\varepsilon$ can be well reproduced by a Gaussian with zero mean and a unity standard deviation when we adopt $\gamma = 0.66$ (see Fig.~\ref{BIS_error_linear}). 

We conclude that the procedure introduced in Sect.~\ref{indic_def_err} overestimates the error on the BIS and that the average correction factor to be adopted to evaluate the true standard deviation is $0.44 < \gamma < 0.66$. 
However, we do not introduce this correction factor in our procedure because we prefer to overestimate the errors on the BIS rather than underestimate them, given that  $\gamma$ is an average factor obtained from a sample including different stars. If required, in the case of a specific star having a homogenous and well-sampled dataset, the above  statistical analysis can be repeated to correct a posteriori the BIS error obtained with our procedure. 

\section{Discussion and conclusions}
We explored the performance of different asymmetry indicators of the CCF, its FWHM, and the chromospheric index $\log R^{\prime}_{\rm HK}$ to detect activity-related RV variations in a sample of 15 slowly rotating ($v \sin i \la 5$ km\,s$^{-1}$), low-activity ($\log R^{\prime}_{\rm HK} \la -4.95$) solar-like stars observed with HARPS-N for two years. We gave prescriptions to compute CCF indicators and their errors starting from the CCFs provided by the DRS of HARPS-N. After removing the RV modulations produced by known planets, we looked for periodic signals in the residuals finding no significant case for additional planets in our sample. Therefore, we computed a kernel regression (KR) of the residual RV with respect to the time and each of our activity indicators,  including the newly defined $V_{\rm asy (mod)}$,  detecting at least one significant correlation (two-sided p-value $<  0.01$) in  seven out of 15 stars. This gives an estimate of the global performance of the KR with our five indicators that was successful in $\sim 47 \pm 18$ percent of the cases.  The bisector inverse slope (BIS), the FWHM, and the new $V_{\rm asy(mod)}$  proved to be the most useful indicators because they provided a significant regression in four, two, and two cases out of 15, respectively. 

In addition to the slow stellar rotation and low activity level,  our relatively low performances could be a consequence of the  conservative significance threshold ($\alpha < 0.01$) chosen  to minimize the occurrence of spurious correlations in our rather large sample. The performance of the CCF asymmetry indicators  improves if we consider stars with an higher activity level or $v\sin i$ such as HD~189733 or $\tau$~Bootis. With $v \sin i = 2.97 \pm 0.2$ km~s$^{-1}$ and $\log R^{\prime}_{\rm HK}$  ranging from $-4.524$ and $-4.458$ \citep{Pace13}, HD~189733 shows a strong correlation between the BIS and the RV residuals with a false-alarm probability lower than 0.1 percent \citep{Boisseetal09}. A similarly strong correlation has been detected in $\tau$~Boo that has $v \sin i = 14.27 \pm 0.06$ km~s$^{-1}$ and $\log R^{\prime}_{\rm HK}$ ranging from $-4.790$ and $-4.768$ in the observations performed by \citet{Borsaetal15}.   
 Our results demonstrate that  KR can be fruitfully applied to our datasets that are characterized by a rather sparse sampling. Kernel regression offers advantages over simple linear or rank correlation analyses because it is not limited to assume a linear or monotone relationship between the RV variations and the activity indicators.  

We note that KR uses two bandwidths that are estimated by minimizing the sum of the squared residuals of the regression (cf.~Sect.~\ref{correlation_coefficients}).  They account for the evolution timescale of the surface active regions and the non-linear dependence of the RV perturbation on the activity indicators. In the implementation presented in this work, they are held fixed because we considered a time span of about two years, but it is simple to introduce a  variation with  the time to account for systematic changes along a stellar activity cycle with a typical duration of several years.  Another advantage of the KR approach is the possibility of analytically estimating the significance of the regression by means of Fisher-Snedecor statistics. 

A consistent comparison of  KR with other techniques requires its inclusion in a global model of the RV variations taking into account both the keplerian motions and the effects of activity and allowing a Bayesian estimate of  the most probable number of planets and of their orbital parameters \citep[cf.][]{Dumusqueetal17}. This is outside the scope of the present paper, where we considered the evaluation of different asymmetry indicators of the CCF and adopted a simple  approach to demonstrate the performance of KR with respect to different activity indicators considering a rather large sample of sun-like stars. The inclusion of KR into a complete RV model will be presented and discussed in future works dedicated to the modelling of the planetary systems of specific targets. The present results suggest that KR can account for a large part of the RV variations induced by stellar activity as shown by the remarkable reduction of the  RV variability when we use KR with our suite of activity indicators. Our approach leads to a reduction of the standard deviations of the RV residuals  that can exceed a factor of approximately two in the case of  significant correlations  (cf. Tables~\ref{results_rhkcorr} and~\ref{results_no_rhkcorr}). 

The best evaluation of the indicators discussed in the present work requires a high-resolution and a stabilized spectrograph such as HARPS-N. A spectral resolution allowing $\lambda/\Delta \lambda$ of at least $10^{5}$ is recommended  in the case of our slowly rotating stars as pointed out for example by \citet{Dumusqueetal14}. With a lower resolution, the evaluation of the indicators is not optimal because the rotational broadening of the spectral lines cannot be resolved in  stars with $v \sin i \la 2-3$~km\,s$^{-1}$. The long-term stability of the spectrograph is another fundamental issue because it  warrants that the observed variation of the FWHM is not dominated by instrumental effects. Actually, we saw the remarkable variation of the FWHM produced by a slightly variable defocussing of HARPS-N during the initial phase of its operation \citep[see][]{Benattietal17}. On the other hand, the impact on the asymmetry indicators was found to be negligible in comparison with the errors associated with the photon shot noise at least in HARPS-N. If its behaviour can be used as a guideline for similar high-resolution spectrographs, this result suggests asymmetry indicators can be used instead of the FWHM when a high stability of the spectrograph is not warranted. 
\begin{acknowledgements}
The authors wish to thank the anonymous referee whose valuable comments on a previous version of this work allowed them to improve the methodology of data analysis and the results. The present work is based on observations made with the high-resolution spectrograph HARPS-N at the Italian {\it Telescopio Nazionale Galileo} (TNG). Data have been made available through the INAF-TNG~IA2 national archive at INAF-Osservatorio astronomico di Trieste whose availability is grateful acknowledged. Simulations obtained with SOAP~2.0 through its web interface \citep{Dumusqueetal14} are gratefully acknowledged. 
The research leading to these results received funding from the European Union Seventh Framework Programme (FP7/2007- 2013) under grant agreement number 313014 (ETAEARTH). 
Support from the {\it Progetti Premiali} scheme (Premiale WOW) of the Italian national Ministry of Education, University, and Research is also acknowledged. PF acknowledges support by Funda\c{c}\~ao para a Ci\^encia e a Tecnologia (FCT) through Investigador FCT contracts of reference IF/01037/2013/CP1191/CT0001, respectively, and POPH/FSE (EC) by FEDER funding through the programme "Programa Operacional de Factores de Competitividade - COMPETE''. PF further acknowledges support from FCT in the form of an exploratory project of reference IF/01037/2013/CP1191/CT0001. The authors are grateful to Dr. R. Zanmar Sanchez for his comments on the present work. 
\end{acknowledgements}
\onecolumn
\begin{longtable}{rccccccccc}
\caption{\label{results_rhkcorr_all}Results of the KR of the RV residuals of the keplerian model including a linear correlation between 
the RV and the stellar chromospheric index $\log R^{\prime}_{\rm HK}$ for our stars. All the analysed cases are listed.} \\
\hline
                                     Star name & Indicator & $N_{\rm c}$ & $\sigma$ & $\sigma_{\rm KR\, res}$ & $h_{\rm t}$ & $h_{\rm x}$ & $\nu$ & $F$ & $\alpha$ \\
                                                      & & & (m/s) & (m/s) & (d) & & & & \\
\hline                                                      
     HD11506 &   $\log R^{\prime}_{\rm HK}$ &     35 &        3.259 &        1.830 &       60.795 &    1.039e-02 &       18.821 &        1.844 &     0.117289 \\
     HD11506 &                          BIS &     35 &        3.029 &        2.069 &       64.575 &    1.819e-03 &       16.401 &        1.307 &     0.292464 \\
     HD11506 &                   $\Delta V$ &     36 &        3.410 &        2.619 &       60.795 &    1.806e-03 &       15.807 &        0.901 &     0.574305 \\
     HD11506 &          $V_{\rm asy (mod)}$ &     36 &        3.410 &        2.248 &       60.987 &    1.797e-03 &       15.598 &        1.728 &     0.129194 \\
     HD11506 &                         FWHM &     36 &        3.410 &        3.013 &       64.575 &    4.251e-03 &       19.901 &        0.224 &     0.998601 \\
                                                          & & & & & & & & &  \\
     HD13931 &   $\log R^{\prime}_{\rm HK}$ &     65 &        2.078 &        1.290 &       35.976 &    7.675e-03 &       38.102 &        1.110 &     0.395993 \\
     HD13931 &                          BIS &     66 &        2.162 &        1.732 &       55.469 &    1.225e-03 &       40.039 &        0.356 &     0.998111 \\
     HD13931 &                   $\Delta V$ &     65 &        2.078 &        1.623 &       73.077 &    1.301e-03 &       43.529 &        0.306 &     0.999443 \\
     HD13931 &          $V_{\rm asy (mod)}$ &     65 &        2.078 &        1.350 &       36.830 &    2.077e-03 &       36.188 &        1.082 &     0.419110 \\
     HD13931 &                         FWHM &     64 &        2.038 &        1.205 &       54.810 &    1.124e-03 &       33.782 &        1.656 &     0.085422 \\
                                                          & & & & & & & & &  \\
     HD23596 &   $\log R^{\prime}_{\rm HK}$ &     25 &        2.642 &        1.634 &       85.161 &    1.147e-02 &       11.541 &        1.902 &     0.140002 \\
     HD23596 &                          BIS &     25 &        2.642 &        1.445 &       85.161 &    1.655e-03 &        8.879 &        4.500 &     0.005952 \\
     HD23596 &                   $\Delta V$ &     25 &        2.642 &        1.347 &       85.161 &    1.931e-03 &       10.182 &        4.269 &     0.007890 \\
     HD23596 &          $V_{\rm asy (mod)}$ &     25 &        2.642 &        1.593 &       85.161 &    2.459e-03 &       11.679 &        2.015 &     0.120641 \\
     HD23596 &                         FWHM &     25 &        2.642 &        1.411 &       85.161 &    1.815e-03 &        8.284 &        5.401 &     0.002504 \\
                                                          & & & & & & & & &  \\
     HD72659 &   $\log R^{\prime}_{\rm HK}$ &     22 &        2.185 &        1.707 &      100.652 &    1.579e-02 &       11.650 &        0.560 &     0.817337 \\
     HD72659 &                          BIS &     22 &        2.185 &        1.515 &      100.652 &    1.682e-03 &       10.215 &        1.261 &     0.354363 \\
     HD72659 &                   $\Delta V$ &     22 &        2.185 &        1.542 &      100.652 &    2.008e-03 &       11.073 &        0.992 &     0.505300 \\
     HD72659 &          $V_{\rm asy (mod)}$ &     21 &        1.872 &        1.057 &      103.492 &    2.300e-03 &        9.506 &        2.622 &     0.072255 \\
     HD72659 &                         FWHM &     22 &        2.185 &        1.387 &      100.652 &    1.805e-03 &        9.069 &        2.167 &     0.109939 \\
                                                          & & & & & & & & &  \\
     HD73534 &   $\log R^{\prime}_{\rm HK}$ &     29 &        2.529 &        2.071 &       66.081 &    1.521e-02 &       18.024 &        0.288 &     0.988335 \\
     HD73534 &                          BIS &     29 &        2.529 &        2.332 &       66.081 &    2.168e-03 &       19.431 &        0.081 &     0.999995 \\
     HD73534 &                   $\Delta V$ &     29 &        2.529 &        2.034 &       66.081 &    1.966e-03 &       18.170 &        0.311 &     0.983370 \\
     HD73534 &          $V_{\rm asy (mod)}$ &     29 &        2.529 &        1.768 &       66.081 &    1.958e-03 &       16.868 &        0.733 &     0.722080 \\
     HD73534 &                         FWHM &     30 &        3.021 &        2.649 &       64.925 &    2.648e-03 &       19.430 &        0.156 &     0.999656 \\
                                                          & & & & & & & & &  \\
     HD75898 &   $\log R^{\prime}_{\rm HK}$ &     34 &        5.579 &        3.339 &       83.177 &    1.581e-02 &       18.540 &        1.476 &     0.229888 \\
     HD75898 &                          BIS &     34 &        5.579 &        3.260 &       88.375 &    1.463e-03 &       14.972 &        2.486 &     0.035639 \\
     HD75898 &                   $\Delta V$ &     34 &        5.579 &        3.013 &       83.177 &    1.854e-03 &       15.796 &        2.815 &     0.021313 \\
     HD75898 &          $V_{\rm asy (mod)}$ &     34 &        5.579 &        2.914 &       88.375 &    1.568e-03 &       15.792 &        3.102 &     0.013560 \\
     HD75898 &                         FWHM &     34 &        5.579 &        3.330 &       88.375 &    3.025e-03 &       15.609 &        2.151 &     0.064822 \\
                                                          & & & & & & & & &  \\
     HD89307 &   $\log R^{\prime}_{\rm HK}$ &     18 &        2.069 &        1.150 &       87.279 &    3.455e-03 &        7.167 &        3.562 &     0.037719 \\
     HD89307 &                          BIS &     19 &        3.775 &        1.467 &       84.339 &    1.777e-03 &        8.642 &        6.862 &     0.004178 \\
     HD89307 &                   $\Delta V$ &     18 &        2.069 &        1.233 &       87.279 &    1.893e-03 &        8.017 &        2.303 &     0.121534 \\
     HD89307 &          $V_{\rm asy (mod)}$ &     19 &        3.775 &        1.187 &       84.339 &    1.815e-03 &        8.046 &       12.648 &     0.000309 \\
     HD89307 &                         FWHM &     18 &        2.069 &        1.383 &       87.279 &    2.076e-03 &        7.575 &        1.777 &     0.204288 \\
                                                          & & & & & & & & &  \\
     HD99109 &   $\log R^{\prime}_{\rm HK}$ &     26 &        1.477 &        1.180 &       96.610 &    1.238e-02 &       13.245 &        0.542 &     0.849973 \\
     HD99109 &                          BIS &     26 &        1.477 &        1.100 &       96.610 &    6.228e-04 &       11.137 &        1.098 &     0.425789 \\
     HD99109 &                   $\Delta V$ &     26 &        1.477 &        1.101 &       96.610 &    9.996e-04 &       13.631 &        0.715 &     0.718776 \\
     HD99109 &          $V_{\rm asy (mod)}$ &     26 &        1.477 &        0.938 &       96.610 &    1.631e-03 &       12.458 &        1.615 &     0.205904 \\
     HD99109 &                         FWHM &     26 &        1.477 &        1.157 &       96.610 &    1.666e-03 &       13.415 &        0.588 &     0.817291 \\
                                                          & & & & & & & & &  \\
    HD106252 &   $\log R^{\prime}_{\rm HK}$ &     42 &        2.059 &        1.372 &       79.651 &    3.250e-03 &       18.640 &        1.586 &     0.150624 \\
    HD106252 &                          BIS &     43 &        2.350 &        1.259 &       79.651 &    8.857e-04 &       19.965 &        2.884 &     0.009355 \\
    HD106252 &                   $\Delta V$ &     43 &        2.350 &        1.592 &       79.651 &    1.238e-03 &       21.261 &        1.205 &     0.337944 \\
    HD106252 &          $V_{\rm asy (mod)}$ &     42 &        2.059 &        1.385 &       79.651 &    1.368e-03 &       19.952 &        1.342 &     0.255381 \\
    HD106252 &                         FWHM &     43 &        2.350 &        1.861 &       79.651 &    2.102e-03 &       25.605 &        0.396 &     0.981523 \\
                                                          & & & & & & & & &  \\
    HD108874 &   $\log R^{\prime}_{\rm HK}$ &     53 &        2.753 &        2.020 &       42.954 &    1.125e-02 &       31.670 &        0.567 &     0.924236 \\
    HD108874 &                          BIS &     53 &        2.753 &        2.188 &       45.988 &    2.584e-03 &       33.698 &        0.326 &     0.997486 \\
    HD108874 &                   $\Delta V$ &     53 &        2.753 &        2.071 &       45.988 &    1.721e-03 &       31.964 &        0.496 &     0.961261 \\
    HD108874 &          $V_{\rm asy (mod)}$ &     53 &        2.753 &        2.172 &       45.988 &    2.907e-03 &       33.759 &        0.338 &     0.996687 \\
    HD108874 &                         FWHM &     53 &        2.753 &        2.013 &       44.796 &    2.643e-03 &       32.355 &        0.545 &     0.937083 \\
                                                           & & & & & & & & &  \\
    HD155358 &   $\log R^{\prime}_{\rm HK}$ &     43 &        1.947 &        1.089 &       67.353 &    2.200e-03 &       17.987 &        3.095 &     0.005702 \\
    HD155358 &                          BIS &     43 &        1.947 &        1.300 &       84.824 &    1.060e-03 &       25.183 &        0.860 &     0.640833 \\
    HD155358 &                   $\Delta V$ &     43 &        1.947 &        1.206 &       67.353 &    1.247e-03 &       21.861 &        1.552 &     0.164736 \\
    HD155358 &          $V_{\rm asy (mod)}$ &     43 &        1.947 &        1.161 &       67.353 &    1.755e-03 &       23.046 &        1.535 &     0.174650 \\
    HD155358 &                         FWHM &     43 &        1.947 &        1.185 &       67.353 &    2.446e-03 &       24.645 &        1.244 &     0.324205 \\
                                                          & & & & & & & & &  \\
    HD188015 &   $\log R^{\prime}_{\rm HK}$ &     21 &        3.945 &        1.853 &      110.264 &    1.295e-02 &        9.138 &        4.692 &     0.010591 \\
    HD188015 &                          BIS &     21 &        3.945 &        2.731 &      110.264 &    2.504e-03 &        9.717 &        1.281 &     0.348434 \\
    HD188015 &                   $\Delta V$ &     21 &        3.945 &        2.691 &      110.264 &    2.618e-03 &        9.665 &        1.350 &     0.319401 \\
    HD188015 &          $V_{\rm asy (mod)}$ &     21 &        3.945 &        2.745 &      110.264 &    3.112e-03 &        8.315 &        1.680 &     0.206100 \\
    HD188015 &                         FWHM &     21 &        3.945 &        1.885 &      110.264 &    3.183e-03 &        8.384 &        5.300 &     0.006108 \\
                                                          & & & & & & & & &  \\
    HD190228 &   $\log R^{\prime}_{\rm HK}$ &     42 &        2.617 &        1.802 &       55.243 &    3.790e-03 &       19.834 &        1.240 &     0.314652 \\
    HD190228 &                          BIS &     43 &        2.621 &        1.638 &       55.243 &    7.001e-04 &       20.490 &        1.722 &     0.111250 \\
    HD190228 &                   $\Delta V$ &     43 &        2.621 &        1.550 &       43.463 &    5.622e-04 &       19.390 &        2.286 &     0.031824 \\
    HD190228 &          $V_{\rm asy (mod)}$ &     43 &        2.621 &        1.437 &       41.111 &    8.937e-04 &       17.415 &        3.482 &     0.002628 \\
    HD190228 &                         FWHM &     43 &        2.621 &        1.975 &       55.243 &    1.361e-03 &       24.318 &        0.576 &     0.894421 \\
                                                          & & & & & & & & &  \\
    HD220773 &   $\log R^{\prime}_{\rm HK}$ &     45 &        2.711 &        1.734 &       58.780 &    9.625e-03 &       27.810 &        0.869 &     0.637097 \\
    HD220773 &                          BIS &     45 &        2.711 &        1.755 &       50.898 &    1.695e-03 &       22.501 &        1.385 &     0.228059 \\
    HD220773 &                   $\Delta V$ &     45 &        2.711 &        1.960 &       61.237 &    1.678e-03 &       25.111 &        0.715 &     0.783846 \\
    HD220773 &          $V_{\rm asy (mod)}$ &     45 &        2.711 &        1.959 &       60.356 &    1.790e-03 &       23.191 &        0.858 &     0.638449 \\
    HD220773 &                         FWHM &     45 &        2.711 &        1.845 &       58.780 &    1.672e-03 &       22.484 &        1.161 &     0.365886 \\
                                                          & & & & & & & & &  \\
        XO2S &   $\log R^{\prime}_{\rm HK}$ &     43 &        3.688 &        2.174 &       79.158 &    4.208e-02 &       21.007 &        1.964 &     0.066427 \\
        XO2S &                          BIS &     42 &        3.534 &        1.513 &       70.129 &    3.774e-03 &       19.207 &        5.312 &     0.000171 \\
        XO2S &                   $\Delta V$ &     42 &        3.534 &        1.925 &       85.365 &    5.877e-03 &       25.640 &        1.449 &     0.227433 \\
        XO2S &          $V_{\rm asy (mod)}$ &     43 &        3.688 &        2.077 &       79.158 &    8.342e-03 &       25.010 &        1.514 &     0.191070 \\
        XO2S &                         FWHM &     42 &        3.341 &        2.108 &       79.158 &    5.697e-03 &       21.635 &        1.409 &     0.226902 \\
\hline
\end{longtable}

\begin{longtable}{rccccccccc}
\caption{\label{results_no_rhkcorr_all}Results of the KR of the RV residuals of the keplerian model that does not include the linear correlation between  
the RV and the  stellar chromospheric index $\log R^{\prime}_{\rm HK}$ for our stars. All the analysed cases are listed. }\\
                    \hline
                                     Star name & Indicator & $N_{\rm c}$ & $\sigma$ & $\sigma_{\rm KR\, res}$ & $h_{\rm t}$ & $h_{\rm x}$ & $\nu$ & $F$ & $\alpha$ \\
                                                      & & & (m/s) & (m/s) & (d) & & & & \\
                     \hline 
     HD11506 &   $\log R^{\prime}_{\rm HK}$ &     35 &        3.232 &        1.858 &       60.795 &    1.039e-02 &       18.821 &        1.723 &     0.145196 \\
     HD11506 &                          BIS &     35 &        3.420 &        2.160 &       61.868 &    1.819e-03 &       16.104 &        1.785 &     0.120970 \\
     HD11506 &                   $\Delta V$ &     36 &        3.566 &        2.767 &       61.868 &    1.865e-03 &       16.234 &        0.815 &     0.653326 \\
     HD11506 &          $V_{\rm asy (mod)}$ &     36 &        3.566 &        2.293 &       64.575 &    1.797e-03 &       16.070 &        1.780 &     0.117630 \\
     HD11506 &                         FWHM &     36 &        3.566 &        3.224 &       64.575 &    4.251e-03 &       19.901 &        0.179 &     0.999673 \\
                                                          & & & & & & & & &  \\
     HD13931 &   $\log R^{\prime}_{\rm HK}$ &     65 &        2.076 &        1.291 &       35.976 &    7.675e-03 &       38.102 &        1.106 &     0.399654 \\
     HD13931 &                          BIS &     66 &        2.155 &        1.721 &       55.469 &    1.225e-03 &       40.039 &        0.363 &     0.997774 \\
     HD13931 &                   $\Delta V$ &     65 &        2.076 &        1.615 &       73.077 &    1.301e-03 &       43.529 &        0.313 &     0.999319 \\
     HD13931 &          $V_{\rm asy (mod)}$ &     65 &        2.076 &        1.343 &       36.463 &    2.077e-03 &       36.084 &        1.106 &     0.396082 \\
     HD13931 &                         FWHM &     65 &        2.076 &        1.347 &       54.516 &    1.124e-03 &       34.775 &        1.188 &     0.319579 \\
                                                          & & & & & & & & &  \\
     HD23596 &   $\log R^{\prime}_{\rm HK}$ &     25 &        3.384 &        1.625 &       85.161 &    1.147e-02 &       11.541 &        3.935 &     0.012499 \\
     HD23596 &                          BIS &     25 &        3.384 &        1.636 &       85.161 &    1.655e-03 &        8.879 &        6.289 &     0.001154 \\
     HD23596 &                   $\Delta V$ &     25 &        3.384 &        1.606 &       85.161 &    1.931e-03 &       10.182 &        5.176 &     0.003308 \\
     HD23596 &          $V_{\rm asy (mod)}$ &     25 &        3.384 &        1.556 &       85.161 &    2.459e-03 &       11.679 &        4.303 &     0.008842 \\
     HD23596 &                         FWHM &     25 &        3.384 &        1.665 &       85.161 &    1.815e-03 &        8.284 &        6.748 &     0.000786 \\
                                                          & & & & & & & & &  \\
     HD72659 &   $\log R^{\prime}_{\rm HK}$ &     22 &        2.260 &        1.709 &      100.652 &    1.579e-02 &       11.650 &        0.657 &     0.745763 \\
     HD72659 &                          BIS &     22 &        2.260 &        1.440 &      100.652 &    1.682e-03 &       10.215 &        1.710 &     0.199798 \\
     HD72659 &                   $\Delta V$ &     22 &        2.260 &        1.480 &      100.652 &    2.008e-03 &       11.073 &        1.311 &     0.339110 \\
     HD72659 &          $V_{\rm asy (mod)}$ &     22 &        2.260 &        1.332 &      100.652 &    2.237e-03 &       10.324 &        2.138 &     0.118953 \\
     HD72659 &                         FWHM &     22 &        2.260 &        1.357 &      100.652 &    1.805e-03 &        9.069 &        2.579 &     0.067751 \\
                                                          & & & & & & & & &  \\
     HD73534 &   $\log R^{\prime}_{\rm HK}$ &     29 &        2.622 &        2.080 &       66.081 &    1.521e-02 &       18.024 &        0.344 &     0.974384 \\
     HD73534 &                          BIS &     29 &        2.622 &        2.343 &       66.081 &    2.168e-03 &       19.431 &        0.117 &     0.999935 \\
     HD73534 &                   $\Delta V$ &     29 &        2.622 &        2.029 &       66.081 &    1.966e-03 &       18.170 &        0.381 &     0.961495 \\
     HD73534 &          $V_{\rm asy (mod)}$ &     29 &        2.622 &        1.887 &       66.081 &    1.958e-03 &       16.868 &        0.651 &     0.788174 \\
     HD73534 &                         FWHM &     30 &        3.084 &        2.642 &       64.925 &    2.648e-03 &       19.430 &        0.188 &     0.998920 \\
                                                          & & & & & & & & &  \\
     HD75898 &   $\log R^{\prime}_{\rm HK}$ &     34 &        5.734 &        3.285 &       83.177 &    1.581e-02 &       18.540 &        1.685 &     0.160382 \\
     HD75898 &                          BIS &     34 &        5.734 &        3.364 &       88.375 &    1.463e-03 &       14.972 &        2.457 &     0.037455 \\
     HD75898 &                   $\Delta V$ &     34 &        5.734 &        3.007 &       85.776 &    1.854e-03 &       15.953 &        2.997 &     0.016108 \\
     HD75898 &          $V_{\rm asy (mod)}$ &     34 &        5.734 &        3.069 &       88.375 &    1.568e-03 &       15.792 &        2.897 &     0.018690 \\
     HD75898 &                         FWHM &     34 &        5.734 &        3.420 &       88.375 &    3.025e-03 &       15.609 &        2.155 &     0.064382 \\
                                                          & & & & & & & & &  \\
     HD89307 &   $\log R^{\prime}_{\rm HK}$ &     18 &        2.034 &        1.150 &       87.279 &    3.455e-03 &        7.167 &        3.387 &     0.043602 \\
     HD89307 &                          BIS &     19 &        3.765 &        1.446 &       84.339 &    1.777e-03 &        8.642 &        7.064 &     0.003759 \\
     HD89307 &                   $\Delta V$ &     18 &        2.034 &        1.189 &       87.279 &    1.893e-03 &        8.017 &        2.439 &     0.106561 \\
     HD89307 &          $V_{\rm asy (mod)}$ &     19 &        3.765 &        1.187 &       84.339 &    1.815e-03 &        8.046 &       12.582 &     0.000316 \\
     HD89307 &                         FWHM &     18 &        2.034 &        1.346 &       87.279 &    2.076e-03 &        7.575 &        1.837 &     0.191571 \\
                                                          & & & & & & & & &  \\
     HD99109 &   $\log R^{\prime}_{\rm HK}$ &     26 &        1.474 &        1.183 &       96.610 &    1.238e-02 &       13.245 &        0.531 &     0.857862 \\
     HD99109 &                          BIS &     26 &        1.474 &        1.075 &       96.610 &    6.228e-04 &       11.137 &        1.203 &     0.366102 \\
     HD99109 &                   $\Delta V$ &     26 &        1.474 &        1.110 &       96.610 &    9.996e-04 &       13.631 &        0.684 &     0.743500 \\
     HD99109 &          $V_{\rm asy (mod)}$ &     26 &        1.474 &        0.957 &       96.610 &    1.631e-03 &       12.458 &        1.506 &     0.241376 \\
     HD99109 &                         FWHM &     26 &        1.474 &        1.165 &       96.610 &    1.666e-03 &       13.415 &        0.562 &     0.836585 \\
                                                          & & & & & & & & &  \\
    HD106252 &   $\log R^{\prime}_{\rm HK}$ &     42 &       2.092 &        1.373 &       79.651 &    3.250e-03 &       18.640 &        1.676 &     0.124049 \\
    HD106252 &                          BIS &     43 &        2.365 &        1.269 &       79.651 &    8.857e-04 &       19.965 &        2.871 &     0.009616 \\
    HD106252 &                   $\Delta V$ &     43 &        2.365 &        1.622 &       79.651 &    1.238e-03 &       21.261 &        1.153 &     0.374282 \\
    HD106252 &          $V_{\rm asy (mod)}$ &     42 &        2.092 &        1.441 &       79.651 &    1.368e-03 &       19.952 &        1.230 &     0.320981 \\
    HD106252 &                         FWHM &     43 &        2.365 &        1.895 &       79.651 &    2.102e-03 &       25.605 &        0.371 &     0.987142 \\
                                                          & & & & & & & & &  \\
    HD108874 &   $\log R^{\prime}_{\rm HK}$ &     53 &        2.753 &        2.020 &       42.954 &    1.125e-02 &       31.670 &        0.567 &     0.924207 \\
    HD108874 &                          BIS &     53 &        2.753 &        2.188 &       45.988 &    2.584e-03 &       33.698 &        0.325 &     0.997521 \\
    HD108874 &                   $\Delta V$ &     53 &        2.753 &        2.073 &       45.988 &    1.721e-03 &       31.964 &        0.495 &     0.961944 \\
    HD108874 &          $V_{\rm asy (mod)}$ &     53 &        2.753 &        2.174 &       45.988 &    2.907e-03 &       33.759 &        0.336 &     0.996791 \\
    HD108874 &                         FWHM &     53 &        2.753 &        2.011 &       44.796 &    2.643e-03 &       32.355 &        0.547 &     0.935833 \\
                                                          & & & & & & & & &  \\
    HD155358 &   $\log R^{\prime}_{\rm HK}$ &     43 &        1.976 &        1.149 &       78.929 &    2.200e-03 &       19.361 &        2.405 &     0.024657 \\
    HD155358 &                          BIS &     43 &        1.976 &        1.328 &       84.824 &    1.060e-03 &       25.183 &        0.841 &     0.658860 \\
    HD155358 &                   $\Delta V$ &     43 &        1.976 &        1.201 &       67.353 &    1.247e-03 &       21.861 &        1.646 &     0.134703 \\
    HD155358 &          $V_{\rm asy (mod)}$ &     43 &        1.976 &        1.171 &       67.353 &    1.755e-03 &       23.046 &        1.570 &     0.162185 \\
    HD155358 &                         FWHM &     43 &        1.976 &        1.201 &       67.353 &    2.446e-03 &       24.645 &        1.248 &     0.321329 \\
                                                          & & & & & & & & &  \\
    HD188015 &   $\log R^{\prime}_{\rm HK}$ &     21 &        4.409 &        1.724 &      110.264 &    1.295e-02 &        9.138 &        7.392 &     0.001748 \\
    HD188015 &                          BIS &     21 &        4.409 &        3.361 &      110.264 &    2.504e-03 &        9.717 &        0.848 &     0.590292 \\
    HD188015 &                   $\Delta V$ &     21 &        4.409 &        3.342 &      110.264 &    2.618e-03 &        9.665 &        0.876 &     0.570951 \\
    HD188015 &          $V_{\rm asy (mod)}$ &     21 &        4.409 &        3.256 &      110.264 &    3.112e-03 &        8.315 &        1.323 &     0.321135 \\
    HD188015 &                         FWHM &     21 &        4.409 &        1.902 &      110.264 &    3.183e-03 &        8.384 &        6.868 &     0.002084 \\
                                                          & & & & & & & & &  \\
    HD190228 &   $\log R^{\prime}_{\rm HK}$ &     42 &        2.614 &        1.802 &       55.243 &    3.790e-03 &       19.833 &        1.235 &     0.317631 \\
    HD190228 &                          BIS &     43 &        2.619 &        1.636 &       55.243 &    7.001e-04 &       20.490 &        1.722 &     0.111434 \\
    HD190228 &                   $\Delta V$ &     43 &        2.619 &        1.548 &       43.463 &    5.598e-04 &       19.352 &        2.296 &     0.031121 \\
    HD190228 &          $V_{\rm asy (mod)}$ &     43 &        2.619 &        1.435 &       41.111 &    8.937e-04 &       17.415 &        3.486 &     0.002610 \\
    HD190228 &                         FWHM &     43 &        2.619 &        1.976 &       55.243 &    1.361e-03 &       24.318 &        0.573 &     0.896383 \\
                                                          & & & & & & & & &  \\
    HD220773 &   $\log R^{\prime}_{\rm HK}$ &     45 &        2.711 &        1.734 &       58.780 &    9.625e-03 &       27.810 &        0.869 &     0.637097 \\
    HD220773 &                          BIS &     45 &        2.711 &        1.755 &       50.898 &    1.695e-03 &       22.501 &        1.385 &     0.228059 \\
    HD220773 &                   $\Delta V$ &     45 &        2.711 &        1.960 &       61.237 &    1.678e-03 &       25.111 &        0.715 &     0.783846 \\
    HD220773 &          $V_{\rm asy (mod)}$ &     45 &        2.711 &        1.959 &       60.356 &    1.790e-03 &       23.191 &        0.858 &     0.638449 \\
    HD220773 &                         FWHM &     45 &        2.711 &        1.845 &       58.780 &    1.672e-03 &       22.484 &        1.161 &     0.365886 \\
                                                          & & & & & & & & &  \\
        XO2S &   $\log R^{\prime}_{\rm HK}$ &     43 &        3.680 &        2.170 &       79.158 &    4.208e-02 &       21.007 &        1.962 &     0.066702 \\
        XO2S &                          BIS &     42 &        3.538 &        1.537 &       72.829 &    3.774e-03 &       19.499 &        4.969 &     0.000292 \\
        XO2S &                   $\Delta V$ &     42 &        3.538 &        1.958 &       85.365 &    5.877e-03 &       25.640 &        1.387 &     0.255739 \\
        XO2S &          $V_{\rm asy (mod)}$ &     43 &        3.680 &        2.077 &       79.158 &    8.341e-03 &       25.009 &        1.504 &     0.194794 \\
        XO2S &                         FWHM &     42 &        3.296 &        2.124 &       79.158 &    5.835e-03 &       21.849 &        1.285 &     0.292560 \\                    
\hline
\end{longtable}
\twocolumn

\appendix
\section{An IDL procedure to compute the asymmetry indicators}
\label{appendix}
We provide a procedure written in IDL 8.4 to compute the CCF asymmetry indicators and FWHM according to the methods described in Sect.~\ref{methods}. It provides also the old indicator $V_{\rm asy}$ as defined by \citet{Figueiraetal13} for reference to previous work, although it is not recommended due to its tendency to show spurious correlations with the RV variations. 

To compile our procedure, installation of the IDL Astronomy Library is required together with the IDL procedure mpfit.pro (we used version 1.82) that can be downloaded from \url{http://cow.physics.wisc.edu/~craigm/idl/fitting.html} with its documentation. You will also need readcol.pro to read ASCII input files and readfits.pro that we included into our file for simplicity, although they are generally found in the IDL Astronomy Library. 

The input data to our procedure consist of a set of fits files produced by the HARPS data reduction software (DRS), specifically those containing the CCF and the bisector profiles, the names of which are of the kind HARPN.YYYY-MM-DDTHH-MM-SS.SSS\_ccf\_MASK\_A.fit and HARPN.YYYY-MM-DDTHH-MM-SS.SSS\_bis\_MASK\_A.fit, respectively. The naming convention for indicating the reduced data files of the DRS is introduced in the HARPS-N User Manual (Sect. 8) and in the DRS User Manual that are accessible through the web page: \url{http://www.tng.iac.es/instruments/harps/}. We recommend reading those manuals before using our IDL procedure. The bisector files are used only for the purpose of comparing our bisector profiles with those provided by the DRS. 

The output of our procedure is an ASCII file giving for each of the input *\_ccf\_* and *\_bis\_* files the asymmetry indicators, the FWHM, and the contrast (central depth relative to the continuum) of the CCF together with their uncertainties. The correction factor $\gamma$ for the BIS error, introduced in the final part of Sect.~\ref{results}, is not included. Our procedure provides the standard deviations of  $\Delta V$ and FWHM computed with the method described in Sect.~\ref{methods} together with those obtained from the covariance matrix and the best fit residuals given by mpfit.pro, although the latter are not recommended for a proper evaluation of the uncertainties because they overestimate the errors (cf. Sect.~\ref{methods}). Details on the output can be found in the header comment lines of the procedure itself.

It is possible to use the procedure in an interactive way that allows the user to see screen plots (e.g. on a X11 terminal) of the CCF and its bisector as well as of the fitting functions used to evaluate the indicators. We recommend the analysis of a sequence of CCF profiles first interactively to see the actual contents of the dataset and check the proper fitting of the CCF profiles. After that stage, it is possible to run the procedure automatically to speed up the analysis of the dataset. When some best fits appear to be not completely adequate because of deviations exceeding a typical threshold of 0.1 percent, the procedure warns the user, asking whether to continue or not. In the case of a lack of sufficient information or bad fits, it stops printing a message. We enclose a set of *\_ccf\_* and *\_bis\_* files together with the corresponding output file to test the operation of the procedure. 

A tarfile including the IDL macro and auxiliary files to compile and test its operation can be downloaded from 
the DropBox: \url{https://www.dropbox.com/sh/74z4hb2wksqg9wg/AABak08BU3EjXFKr9XnSVYPva?dl=0} 
or from INAF GitLab at
\url{https://www.ict.inaf.it/gitlab/antonino.lanza/HARPSN_spectral_line_profile_indicators.git} 
\end{document}